\documentclass[aps,prb,twocolumn,floatfix,citeautoscript,hyperref=pdftex]{revtex4-1}
\usepackage{times}
\usepackage{graphicx}
\usepackage{dcolumn}
\usepackage{bm}
\usepackage{color}
\usepackage{amsmath}
\usepackage{amssymb}
\newcommand {\beq} {\begin{equation}}
\newcommand {\eeq} {\end{equation}}
\newcommand {\bqa} {\begin{eqnarray}}
\newcommand {\eqa} {\end{eqnarray}}

\usepackage[colorlinks=true]{hyperref}

\begin{document}

\title{Odd-frequency pair density wave correlations in underdoped cuprates}

\author{Debmalya Chakraborty}
\affiliation{Department of Physics and Astronomy, Uppsala University, Box 516, S-751 20 Uppsala, Sweden}

\author{Annica M. Black-Schaffer}
\affiliation{Department of Physics and Astronomy, Uppsala University, Box 516, S-751 20 Uppsala, Sweden}

\begin{abstract}

Pair density waves, identified by Cooper pairs with finite center-of-mass momentum, have recently been observed in copper oxide based high T$_\textrm{c}$ superconductors (cuprates). A charge density modulation or wave is also ubiquitously found in underdoped cuprates. Within a general mean-field one-band model we show that the coexistence of charge density waves and uniform superconductivity in $d$-wave superconductors like cuprates, generates an odd-frequency spin-singlet pair density wave, in addition to the even-frequency counterparts. The strength of the induced odd-frequency pair density wave depends on the modulation wave vector of the charge density wave, with the odd-frequency pair density waves even becoming comparable to the even-frequency ones in parts of the Brillouin zone. We show that a change in the modulation wave vector of the charge density wave from bi-axial to uni-axial, can enhance the odd-frequency component of the pair density waves. Such a coexistence of superconductivity and uni-axial charge density wave has already been experimentally verified at high magnetic fields in underdoped cuprates. We further discuss the possibility of an odd-frequency spin-triplet pair density wave generated in the coexistence regime of superconductivity and spin density waves, applicable to the iron-based superconductors. Our work thus presents a route to bulk odd-frequency superconductivity in high T$_c$ superconductors.

\end{abstract}

\maketitle

\section{Introduction}\label{sec:Intro}

Broken symmetry phases characterize different condensed matter systems and define their phase diagrams. One of the most coveted phases of matter is superconductivity. Many non-superconducting phases lie in the proximity to superconductivity in various materials, making their phase diagram immensely complex and rich, with charge and spin density waves being two of the primary candidates. The interplay of density waves and superconductivity has already been found in transition-metal dichalcogenides \cite{Joe14,Gye19,Cho18}, twisted bilayer graphene \cite{Isobe18}, twisted double-bilayer graphene \cite{Rickhaus20}, and iron-based \cite{Dai15} and copper oxide based (cuprate) \cite{Keimer15,Fradkin15} superconductors.  

In cuprates, charge density waves (CDW) have been ubiquitously observed in underdoped samples using many experimental probes, such as scanning tunneling microscopy \cite{Hoffman02,Matsuba07,Yoshizawa13,Machida16,Hamidian15a}, x-ray scattering\cite{Chang12,Blanco-Canosa13,Blackburn13a,Ghiringhelli12,Gerber:2015gx,Chang16}, NMR \cite{Wu11,Wu13a,Wu:2015bt,Julien15} and transport measurements \cite{Doiron-Leyraud07,Sebastian12}. CDW have drawn significant attention due to its potential ability \cite{Metlitski10b,Efetov13,Wang14} to explain the mysterious pseudo-gap phase,\cite{Alloul89,Warren89} found at temperatures larger than the superconducting transition temperature T$_\textrm{c}$. It has been argued that CDW compete with superconductivity (SC) in parts of the doping phase diagram \cite{Chang12,Ghiringhelli12}. However, the strength of the competition of CDW and superconductivity clearly varies between different cuprates and a coexistence state is also observed at high magnetic fields \cite{Gerber:2015gx,Chang16,LeBoeuf13,Wu11,Wu13a,Vinograd18}.

Apart from modulations in the charge density, spatial modulations in the superconducting pair amplitude have also been observed using Josephson scanning tunneling microscopy in the cuprate compound Bi$_2$Sr$_2$CaCu$_2$O$_{8+x}$ (BSCCO).\cite{Hamidian16}Modulating superconducting pair amplitude is often referred to as Cooper-pair density waves (PDW). PDW closely resemble the Fulde-Ferrell-Larkin-Ovchinnikov (FFLO) \cite{Fulde64,Larkin64} state, but with no explicit time-reversal symmetry breaking or net magnetization. In contrast to the uniform superconducting state, the spatial average of the superconducting order parameter is zero in the PDW state. The experimental observations of PDW, be it direct \cite{Hamidian16,Ruan18} or indirect \cite{Edkins18}, has always associated it to CDW and a uniform SC. Interestingly, the modulation wave vector of PDW, $Q_{\text{PDW}}$, is either same or half of the CDW modulation wave vector $Q$ \cite{Hamidian16,Edkins18,Ruan18} and the amplitudes of the two modulating orders are directly correlated \cite{Ruan18}.  

The PDW state has also been studied theoretically in the context of striped La-based cuprates \cite{Himeda02, Baruch08, Agterberg08, Berg09, Berg09a} and more recently when explaining the generic phase diagram of cuprates \cite{Wang18,Dai18,Norman18,Choubey20}. Fluctuating PDW have also been proposed to be a candidate for explaining the pseudo-gap phase \cite{Lee14,Chakraborty19,Dai20,Grandadam20}. Many questions regarding the nature of PDW in cuprates are however still left to be answered. For example, it is not yet known whether PDW give rise to CDW \cite{Berg09} or a coexistence of uniform SC and CDW gives rise to PDW \cite{Chen04,Balents05,Melikyan05}. This confusion arises due to multitude of different possibilities \cite{Berg09,Wang18,Dai18}: A primary PDW with $Q_{\text{PDW}}$ can give rise to a secondary CDW with $Q=2Q_{\text{PDW}}$. An additional CDW with $Q=Q_{\text{PDW}}$ arises when the primary PDW coexist with uniform SC. In addition, a coexistence of uniform SC and CDW can induce PDW with $Q_{\text{PDW}}=Q$. Direct measurements of PDW have only observed $Q=Q_{\text{PDW}}$ \cite{Hamidian16,Ruan18}, strengthening the line of thought that the PDW in cuprates are induced due to the coexistence of uniform SC and CDW. However, recent STM experiments observe CDW both at $Q=Q_{\text{PDW}}$ and $Q=2Q_{\text{PDW}}$ \cite{Edkins18}, indirectly indicating that the PDW is a primary order, though CDW with $Q=2Q_{\text{PDW}}$ have not been observed with any other experimental probe. To add to the complexity, a microscopic description of the PDW state itself is also lacking \cite{Agterberg20}.

The PDW order is characterized by finite center-of-mass momentum Cooper pairs and has so far been described by equal time two electron correlation functions, found within conventional BCS-like theory. If instead also unequal time pair correlation functions are considered, i.e. the two electrons can also pair at unequal times, the Fermi-Dirac statistics allows for the exotic possibility that the correlation function becomes odd under the exchange of the time coordinates or, equivalently, odd in frequency \cite{Berezinskii74, Kirkpatrick91, Balatsky92, Schrieffer94, Bergeret05, Linder19}. Odd-frequency Cooper pairs with zero center-of-mass momentum are present in several superconducting systems \cite{Bergeret05,Tanaka12,Linder19,Cayao20,Triola20}, among which most break translational symmetry. The necessary broken translational symmetry has been achieved in junctions \cite{Bergeret01, Tanaka07, Eschrig07, Zhu10, DiBernardo15, Crepin15, Burset15,Cayao18, Hwang18, Lothman20, Parhizgar20, Dutta20, Krieger20}, in the presence of impurities \cite{Triola19, Perrin19, Kuzmanovski20_a, Santos20, Kuzmanovski20}, and also in models with a staggered lattice structure \cite{Wakatsuki14, Kuzmanovski17,Tamura20}. Bulk odd-frequency correlations are also predicted to be present in multiband superconductors even without broken translational symmetry \cite{Black-Schaffer13,Komendova15,Asano15,Komendova17,Triola18,Triola20,Schmidt20}. But, the thermodynamic stability \cite{Heid95} of odd-frequency superconductivity has been questioned due to a most often found paramagnetic, or negative, Meissner response \cite{Hashimoto01,Bergeret01,Bernardo15}. However, a diamagnetic Meissner effect can be restored by having a frequency dependent pair potential \cite{Belitz99,Solenov09,Kusunose11} or if the odd-frequency pairs have a staggered nature \cite{Hoshino14,Hoshino16}, i.e., in other words have finite center-of-mass momentum. Thus odd-frequency superconductivity with a spatially modulated order parameter can be thermodynamically stable. This naturally leads to the tantalizing possibility: is the mysterious PDW state found in the cuprates an odd-frequency state?

In this work, we show that spin-singlet odd-frequency pair density wave (OPDW) correlations are generically found in cuprates due to the simultaneous presence of uniform SC and CDW. We investigate a time-independent Hamiltonian of coexisting CDW and uniform SC, described by equal time BCS-like Cooper pairs with zero center-of-mass momentum, as commonly assumed present in cuprates. We find that this conventional time-independent Hamiltonian induces PDW correlations consisting of unequal time and finite-momentum Cooper pairs, and among those an OPDW. The OPDW in our work is an induced correlation with $Q_{\text{PDW}}=Q$, i.e.~it is not an order parameter and thus does not need an exotic pairing interaction in the OPDW channel. Being induced correlations, the OPDW does not alter the ground state of the Hamiltonian. This makes our work much more general in comparison to the earlier works finding modulated odd-frequency superconductivity in quasi-one-dimensional models \cite{Hoshino14,Hoshino16,Tsvelik19}. We find that the OPDW is also accompanied with an even-frequency pair density wave (EPDW). By exploring various possible CDW modulation wave vectors and different band structures, we show that the OPDW is a generic feature of all cuprates, whereas the relative strength of OPDW and EPDW depends on the specific material. While both OPDW and EPDW are sign changing in momentum space, their momentum structure are characteristically different, giving various possibilities of their experimental identification. We also explore the coexistence of a spin density wave state and SC, and find a spin-triplet OPDW. This triplet OPDW is  unlikely to be found in the cuprates, but might be present in the iron-based superconductors. Our work thus shows a generic pathway to realizing OPDW in several different coexistence phases.

We organize the rest of the article in the following way. In Sec.~\ref{sec:model} we give the details of a general mean-field model of coexisting SC and CDW and discuss the possible OPDW that can be induced. We then turn to the specific case of cuprates in Sec.~\ref{sec:cuprate}. We find explicit values of the induced OPDW by considering three different possible CDW modulation wave vectors in Secs.~\ref{sec:diagonal}-\ref{sec:uniaxial}, before showing in Sec.~\ref{sec:robusttobands} that the OPDW is a robust feature to three different cuprate bands. After that, we investigate the coexistence of SC and spin density waves in Sec.~\ref{sec:SCSDW}. Finally in Sec.~\ref{sec:Discussion}, we summarize our results and discuss various experimental consequences in cuprates and also the relevance of our findings in the context of other materials showing coexistence phases.

\section{Generic Model}\label{sec:model}

We start by considering a general mean-field model of coexisting SC and CDW. The corresponding Hamiltonian in momentum space is given by,
\begin{eqnarray}
H&=&\sum_{k,\sigma} \xi_{k} c_{k \sigma}^{\dagger} c_{k \sigma} + \sum_{k, \sigma} \left( \chi_{k} c_{k \sigma}^{\dagger} c_{k+Q \sigma}+ \textrm{H.c.} \right) \nonumber \\
&+& \sum_{k} \left( \Delta_{k} c_{-k \downarrow} c_{k \uparrow} + \textrm{H.c.} \right)+\text{constant},
\label{eq:Hamil}
\end{eqnarray}
where $c_{k \sigma}^{\dagger}$ ($c_{k \sigma}$) is the creation (annihilation) operator of an electron with spin $\sigma$ and momentum $k$, $\xi_{k}$ is the electron dispersion, $\chi_k$ is the CDW order parameter with a modulation wave vector $Q$ and $\Delta_k$ is the superconducting order parameter. In Eq.~\eqref{eq:Hamil} we have considered a spin-singlet even parity superconducting order parameter, such that $\Delta_{-k}=\Delta_{k}$, and $\chi_k$ also to be parity preserving with $\chi_{-k}=\chi_{k}$. 
Here, $\chi_k=\sum_{k^{\prime}}V^{\chi}_{k,k^{\prime}} \langle c_{k^{\prime} \sigma}^{\dagger} c_{k^{\prime}+Q \sigma} \rangle $ and, likewise, $\Delta_k=\sum_{k^{\prime}}V^{\Delta}_{k,k^{\prime}} \langle c_{k^{\prime} \uparrow}^{\dagger} c_{-k^{\prime} \downarrow}^{\dagger} \rangle$, where $V^{\chi}_{k,k^{\prime}}$ and $V^{\Delta}_{k,k^{\prime}}$ are the interaction strengths driving the CDW and superconducting orders, respectively. The form and values of these interaction strengths depend on the microscopic properties. Here we do not assume any particular microscopic model and simply consider $\chi_k$ and $\Delta_k$ as parameters in order to keep our results as general as possible. 
Note that the notation for the CDW order parameter should ideally be $\chi_{k,k+Q}$, but for brevity we use the short-hand notation $\chi_k$ with the $Q$ index being absorbed in the definition. 

The Hamiltonian in Eq.~\eqref{eq:Hamil} can be written in a matrix form in the basis $\Psi^{\dagger}=\left(c_{k \uparrow}^{\dagger},c_{k+Q \uparrow}^{\dagger},c_{-k \downarrow},c_{-k-Q \downarrow}\right)$ as,
\begin{eqnarray}
&&H=\frac{1}{2}\sum_{k} \Psi^{\dagger} \hat{H} \Psi+\text{constant}, \nonumber \\
&&\text{with} \nonumber \\
&&\hat{H}=\left(\begin{array}{cccc} \xi_k & \chi_{k} & \Delta_{k} & 0 \\
\chi_k & \xi_{k+Q} & 0 & \Delta_{k+Q} \\
\Delta_k & 0 & -\xi_{k} & -\chi_{k} \\
0 & \Delta_{k+Q} & -\chi_{k} & -\xi_{k+Q} \\
\end{array}\right),
\label{eq:Hamilmat}
\end{eqnarray} 
where we take $\chi_k$ and $\Delta_{k}$ to be real, without loss of generality. A purely imaginary $\chi_k$ is often considered to describe current density wave orders \cite{Chakravarty01}. In the rest of this article, we focus only on a purely real $\chi_k$. However, our analysis extends equivalently to a complex $\chi_k$ without any change in the outcomes. The Green's function $G$ corresponding to the Hamiltonian in Eq.~\eqref{eq:Hamilmat} is given by $G^{-1}(i\omega)=i\omega-\hat{H}$ where $\omega$ are fermionic Matsubara frequencies. The Matsubara frequencies are discrete and only become continuous in the zero-temperature limit. As we are focused on the low-temperature physics, we work with continuous Matsubara frequencies in the rest of the article. An alternative and equivalent picture can also be developed by analytically continuing to real frequencies and considering advanced and retarded Green's functions \cite{Linder19}.

\subsection{Induced spin-singlet odd-frequency PDW}\label{sec:OPDW}

In the context of mean-field microscopic models pertaining to cuprates, the search for PDW has sometimes been focused around finding an energetically minimum PDW ground state \cite{Loder10,Waardh17,Waardh18}. However, a PDW ground state is either hard to stabilize energetically or fragile to parameters of the model \cite{Agterberg20}. So, our starting point of this work is simply a mean-field model Hamiltonian (Eq.~\eqref{eq:Hamil}) with only uniform SC and CDW orders, as also considered previously \cite{Chen04, Balents05, Melikyan05}. We thus do not consider any PDW order parameter in the mean-field Hamiltonian and consequently our approach does not require us to stabilize a PDW ground state. Instead, we consider the PDW as induced correlations generated in the model and show that this generated or induced PDW very generally has both even and odd-frequency components. More specifically, this induced PDW can be found by looking at the Cooper pair correlator $F_{k,k+Q}(\tau)=-\langle T_{\tau}c_{k \uparrow}^{\dagger}(\tau) c_{-k-Q \downarrow}^{\dagger}(0) \rangle$, where $\tau$ is the imaginary time and $T_{\tau}$ is the $\tau$-ordering operator. Here $F_{k,k+Q}(\tau)$ has all the symmetries of a PDW field as it describes a Cooper pair with a finite center-of-mass momentum, with the same wave vector $Q$ as that of the CDW. However, note that this induced PDW amplitude given by $F_{k,k+Q}(\tau)$ is a correlation function and is thus not associated with any interaction in the PDW channel. As a result, the PDW correlations do not enter the Hamiltonian and consequently do not alter the free energy of the system. (This is in sharp contrast to e.g.~$F_{k}(\tau)=-\langle T_{\tau}c_{k \uparrow}^{\dagger}(\tau) c_{-k \downarrow}^{\dagger}(0) \rangle$, which in the limit of $\tau=0$ (equal time) and with a finite pairing strength defines the self-consistent gap equation for $\Delta_k$ in the Hamiltonian.)

Using the Hamiltonian in Eq.~\eqref{eq:Hamilmat}, the Green's function is obtained by inverting the $4\times4$ matrix $G^{-1}(i\omega)$. The Fourier transformed PDW correlator is then given by,
\begin{eqnarray}
&&F_{k,k+Q}(i\omega)=G_{14}(i\omega)=F^{even}_{k,k+Q}(i\omega)+F^{odd}_{k,k+Q}(i\omega),\nonumber \\
&&\text{where}\nonumber \\
&&F^{even}_{k,k+Q}(i\omega)=\frac{\chi_{k}\left(\xi_k\Delta_{k+Q}+\xi_{k+Q}\Delta_k\right)}{D}, \label{eq:feven}\\
&&F^{odd}_{k,k+Q}(i\omega)=\frac{i\omega\chi_{k}\left(\Delta_k-\Delta_{k+Q}\right)}{D},\label{eq:fodd} \\
&&D=\left( \omega^2+\xi_{k}^2+\Delta_{k}^2 \right)\left( \omega^2+\xi_{k+Q}^2+\Delta_{k+Q}^2 \right)\nonumber \\
&&-2\chi_{k}^2 \left( \xi_k \xi_{k+Q}-\Delta_{k}\Delta_{k+Q} -\omega^2 \right)+\chi_{k}^4. \label{eq:D}
\end{eqnarray}
The induced PDW, per definition given by $G_{14}(i\omega)$, have both even and odd-frequency components. This is seen explicitly in Eqs.~\eqref{eq:feven} and \eqref{eq:fodd}, with $F^{even}_{k,k+Q}(i\omega)$ and $F^{odd}_{k,k+Q}(i\omega)$ having an even and odd frequency dependence, respectively, in the numerator, while the denominator $D$ is an even function of frequency.

In order to verify the symmetries of $F_{k,k+Q}(\tau)$, we use the fact that a pair correlation function should always satisfy the Fermi-Dirac statistics. As a result, the correlation function under a joint operation of spin permutation (S), momentum exchange (M), and relative time permutation (T) should satisfy $SMT=-1$. Under $M$, $F_{k,k+Q}(\tau)\rightarrow F_{k+Q,k}(\tau)$. So, we look at the $G_{23}(i\omega)$ component of the Green's function, giving,
\begin{eqnarray}
&&F_{k+Q,k}(i\omega)=G_{23}(i\omega)=F^{even}_{k+Q,k}(i\omega)+F^{odd}_{k+Q,k}(i\omega),\nonumber \\
&&\text{where}\nonumber \\
&&F^{even}_{k+Q,k}(i\omega)=\frac{\chi_{k}\left(\xi_k\Delta_{k+Q}+\xi_{k+Q}\Delta_k\right)}{D}=F^{even}_{k,k+Q}(i\omega), \nonumber \\
&&\label{eq:fevenm}\\
&&F^{odd}_{k+Q,k}(i\omega)=\frac{i\omega\chi_{k}\left(\Delta_{k+Q}-\Delta_{k}\right)}{D}=-F^{odd}_{k,k+Q}(i\omega),\label{eq:foddm}
\end{eqnarray}
The even-frequency component of the induced PDW is thus even under $M$ (seen in Eq.~\eqref{eq:fevenm}) and also even under $T$. To satisfy $SMT=-1$, it must therefore be odd under $S$, or a spin-singlet state. Similarly, the odd-frequency component is odd under $M$ (from Eq.~\eqref{eq:foddm}), odd under $T$ and thus, again odd under $S$. As a consequence, both even and odd-frequency components of the induced PDW are spin-singlet in nature, as we also expect since Eq.~\eqref{eq:Hamil} is spin-rotation invariant.

From Eq.~\eqref{eq:fodd}, we can already now gain some insights to the OPDW. If the SC is described by an $s$-wave, or momentum independent, order parameter, the OPDW is zero, since then $\Delta_{k+Q}=\Delta_{k}$ for any $Q$. So, the coexistence of $s$-wave SC and CDW in a single-band system cannot give rise to the OPDW. In contrast, if the superconducting order parameter is momentum dependent, $\Delta_{k+Q}\ne\Delta_{k}$ in general. For example, a coexistence of $d$-wave SC (given by $\Delta_{k}\propto \cos k_x-\cos k_y$) and CDW with $Q=(\pi,\pi)$ gives the highest OPDW as then $\Delta_{k+Q}=-\Delta_{k}$. In the next section, we investigate the OPDW in the context of cuprates, prototype $d$-wave superconductors. Although superconductivity in cuprates is achieved by doping a parent antiferromagnetic ($Q=(\pi,\pi)$) insulator \cite{Lee06}, the CDW observed in these materials have a $Q$ different from $(\pi,\pi)$.

Before discussing the case of cuprates, we comment on the similarities of the Hamiltonian in Eq.~\eqref{eq:Hamilmat} with that of a multiband superconductor. A simple analogy with the Hamiltonian of multiband superconductors can be drawn if we consider $\xi_{k+Q}$ as an independent second band. In this picture, $\chi_k$ is then the band hybridization between two bands $\xi_k$ and $\xi_{k+Q}$, while $\Delta_k$ and $\Delta_{k+Q}$ become two independent superconducting order parameters in each bands, respectively. As was shown in Ref.~[\onlinecite{Black-Schaffer13}], odd-frequency pairing arises if the band hybridization is finite and the superconducting order parameters are not equal in two bands, i.e., if $\Delta_{k+Q}\ne\Delta_{k}$. This is the same criterion as established in Eq.~\eqref{eq:fodd} for CDW and thus illustrates the underlying similarity, although the materials and properties are completely different. We also note that in the case of multiband superconductors, the odd-frequency pairing does not have a modulation wave vector.

\section{Case of cuprates}\label{sec:cuprate}

The origin of CDW in cuprates is still a debated question. Two parallel points of view have been proposed: one based on strong real space electron interactions giving CDW with wave vectors commensurate with the lattice and the other based on a momentum space picture where the Fermi surface plays an important role in defining the CDW wave vectors. The former picture discusses the experimentally observed incommensurate CDW wave vectors by disorder-induced discommensuration effects \cite{Mesaros16}. The latter gives a CDW wave vector that connects points of different branches of the Fermi surface. In this latter picture \cite{Metlitski10b,Efetov13,Wang14}, it has been postulated that there exists an antiferromagnetic quantum critical point beneath the superconducting dome. Outside the antiferromagnetic phase, short-range antiferromagnetic fluctuations diverge near the `hot spots' ($k$-points where the antiferromagnetic Brillouin zone intersects the Fermi surface) in two spatial dimensions, giving rise to CDW and superconducting correlations. As a result, the CDW wave vectors are connecting different `hot spots'. In this work, we consider an effective homogeneous Hamiltonian (Eq.~\eqref{eq:Hamil}) in momentum space where CDW wave vectors are considered within the latter picture, i.e., connecting different `hot spots'. The model in Eq.~\eqref{eq:Hamil} can also describe the former picture with commensurate CDW wave vectors. Therefore the choice of model is not crucial for our results, although we note that disorder-induced discommensuration effects in the first picture cannot be captured within this model.

To continue, we consider in Secs.~\ref{sec:diagonal}-\ref{sec:uniaxial} a band dispersion mimicking the Fermi surface of the underdoped cuprate, YBa$_2$Cu$_3$O$_{7-x}$ (YBCO) \cite{Schabel98} with $\xi_k$ given by,
\begin{eqnarray}
\xi_k&=&\frac{t_1}{2}\left(\cos k_x +\cos k_y\right)+t_2 \cos k_x \cos k_y \nonumber\\
&&+\frac{t_3}{2} \left(\cos 2k_x +\cos 2k_y \right) + \frac{t_4}{2} \left( \cos 2k_x \cos k_y \right.\nonumber \\
&&\left.+\cos k_x \cos 2k_y \right)+t_5 \cos 2k_x \cos 2k_y +\mu,
\label{eq:Band}
\end{eqnarray}
with 
\begin{eqnarray}
\text{YBCO:}~&&t_1=-1.1259~\text{eV}, t_2=0.5540~\text{eV}, t_3=-0.1174~\text{eV}, \nonumber \\
&&t_4=-0.0701~\text{eV}, t_5=0.1286~\text{eV},~\mu=0.1756~\text{eV}. \nonumber \\
\label{eq:Band1parameters}
\end{eqnarray}

It should be noted that the bilayer electronic structure of YBCO results in a splitting of the bands \cite{Schabel98,Torre16}. This gives two sets of band parameters representing the bonding and the antibonding band dispersions. In Eq.~\eqref{eq:Band1parameters}, we consider only the parameters corresponding to the bonding band dispersion. The hole doping is given by the average of the doping in the bonding and the antibonding bands. With this bonding band dispersion, the corresponding hole doping is $12\%$ \cite{Schabel98,Torre16}, which is where the intensity of the CDW in x-ray experiments \cite{Ghiringhelli12} is close to maximum. We will primarily express all energies corresponding to this band in units of $t_1$. Motivated by experiments, superconducting \cite{Scalapino95,Maki98} and CDW \cite{Hamidian15a,Comin15} order parameters are taken to be $d$-wave in nature and given by,
\begin{eqnarray}
\Delta_{k}=\frac{\Delta_{0}}{2} \left( \cos k_x - \cos k_y \right), \label{eq:SCOP} \\
\chi_{k}=\frac{\chi_{0}}{2} \left( \cos k_x - \cos k_y \right), \label{eq:CDWOP}
\end{eqnarray}
where $\Delta_{0}$ and $\chi_0$ gives the maximum values of the superconducting and CDW order parameters, respectively. We note that recent resonant x-ray scattering measurements suggest an unusual possibility of an $s$-wave symmetry of the CDW in YBCO \cite{McMahon19}. Our analysis in Sec.~\ref{sec:OPDW} holds for any symmetry of the CDW order, however the quantitative results in this section will differ with an $s$-wave CDW.    

The exact values of $\Delta_0$ and $\chi_0$ can be calculated self-consistently by choosing a microscopic model with interactions in both the SC and CDW channels. However, these values will crucially depend on the exact choice of microscopic model and, with the microscopic origin of both SC and CDW in underdoped cuprates still being an unresolved question, are at present nearly impossible to determine. In fact, various microscopic models exist giving different phase diagrams and even different relative values of $\Delta_0$ and $\chi_0$ \cite{Metlitski10b,Efetov13,Wang14,Chakraborty19}. So, in order to avoid this ambiguity, we choose a path of instead scanning a wide range of $\Delta_0$ and $\chi_0$, without solving the self-consistent gap equations. This will automatically cover the range of possible self-consistent solution, without hampering our conclusions. Moreover, and most importantly, the ratio of the OPDW and the EPDW for a particular $k$-point in the Brillouin zone and a fixed frequency does not, in fact, depend on the choice of $\Delta_0$ and $\chi_0$. From Eqs.~\eqref{eq:feven} and \eqref{eq:fodd}, and using the form of $\Delta_{k}$ and $\chi_k$ in Eqs.~\eqref{eq:SCOP} and \eqref{eq:CDWOP}, the ratio of the OPDW and the EPDW is given by,
\begin{equation}
\frac{F^{odd}_{k,k+Q}(i\omega)}{F^{even}_{k,k+Q}(i\omega)}=\frac{i\omega \left(\eta_k-\eta_{k+Q}\right)}{\left(\xi_k\eta_{k+Q}+\xi_{k+Q}\eta_k\right)},
\label{eq:selfcon}
\end{equation}
if $\chi_k\ne 0$, $\Delta_0\ne 0$ and $D\ne 0$, and with $\eta_k=\left( \cos k_x - \cos k_y \right)$. Thus, the relative importance of the OPDW to the EPDW correlations is not expected to change with any self-consistent determination of $\Delta_0$ and $\chi_0$, although the quantitative strength of the PDW correlations might of course change.

In order to efficiently study the PDW, we use the fact that the EPDW is even and the OPDW is odd under the $M$ momentum exchange operation, as seen in Eqs.~\eqref{eq:fevenm}-\eqref{eq:foddm}. As a result, the total contribution of the OPDW can be obtained by taking an antisymmetric combination of $F_{k,k+Q}^{odd}$ and $F_{k+Q,k}^{odd}$, while the total EPDW contribution is obtained by taking the symmetric combination of $F_{k,k+Q}^{even}$ and $F_{k+Q,k}^{even}$. Thus, the total contribution of the OPDW and the EPDW are given by,
\begin{eqnarray}
F_{k}^{o}(i\omega)&=&Im\left(\frac{F_{k,k+Q}^{odd}(i\omega)-F_{k+Q,k}^{odd}(i\omega)}{2}\right), \label{eq:OPDW_sym} \\
F_{k}^{e}(i\omega)&=&\frac{F_{k,k+Q}^{even}(i\omega)+F_{k+Q,k}^{even}(i\omega)}{2}, \label{eq:EPDW_sym}
\end{eqnarray} 
where we have taken the imaginary part in the first line as both $F_{k,k+Q}^{odd}(i\omega)$ and $F_{k+Q,k}^{odd}(i\omega)$ are purely imaginary, see Eq.~\eqref{eq:foddm}. 
Looking at the functional form of the EPDW and OPDW correlations in Eqs.~\eqref{eq:feven} and \eqref{eq:fodd}, it is evident that the momentum structure of $F_{k}^{e}$ and $F_{k}^{o}$ depend on the $Q$ vector. In addition, we consider both CDW and SC to have $d$-wave symmetry. As the induced PDW comes as a product of CDW and SC in Eqs.~\eqref{eq:feven} and \eqref{eq:fodd}, we can already now conclude that $F_{k}^{e}$ and $F_{k}^{o}$ do not have a simple $d$-wave structure. Still, we expect sign changes in $F_{k}^{e}$ and $F_{k}^{o}$ as they depend on both $k$ and $k+Q$. Keeping in mind that we might encounter  sign changes in the induced PDW, we define the following two momentum sums to characterize the total momentum contributions,
\begin{eqnarray}
F^{e/o}(i\omega)&=&\sum_{k} \left| F_{k}^{e/o}(i\omega) \right|, \label{eq:EOPDW_modksum} \\
F^{e/o}_{*}(i\omega)&=&\sum_{k} F_{k}^{e/o}(i\omega). \label{eq:EOPDW_ksum}
\end{eqnarray}
If $F_{k}^{e/o}$ has a pure $d$-wave structure, $F^{e/o}_{*}$ will be zero. 

The Fourier transformed PDW correlators $F_{k}^{o}(i\omega)$ and $F_{k}^{e}(i\omega)$ encode all the properties of the superconducting pairing in the system and are thus the key quantities for providing any deeper understanding of the superconducting phase.
They also play a crucial role in the determination of two-particle response functions, which can then be measured experimentally. For example, density response functions can be measured experimentally in various probes like resonant inelastic x-ray scattering (RIXS), momentum-resolved electron-loss spectroscopy (M-EELS), and Raman spectroscopy. Moreover, one of the hallmarks of superconductivity, the Meissner response, is a current response function and is also affected by the PDW correlators. The Meissner response has previously been instrumental in detecting the signatures of odd-frequency superconductivity, when giving a paramagnetic response \cite{Hashimoto01,Bergeret01,Bernardo15} in contrast to the usual diamagnetic response of even-frequency superconductivity. Depending on the experimental probes, either the momentum dependent correlators or momentum averaged correlators can be relevant. Thus, in the next sections, we calculate both the momentum structure and momentum averaged quantities for the induced PDW, choosing three specific $Q$ values discussed in the context of cuprates.

\subsection{Diagonal CDW}\label{sec:diagonal}

\begin{figure}[htb]
\includegraphics[width=1.0\linewidth]{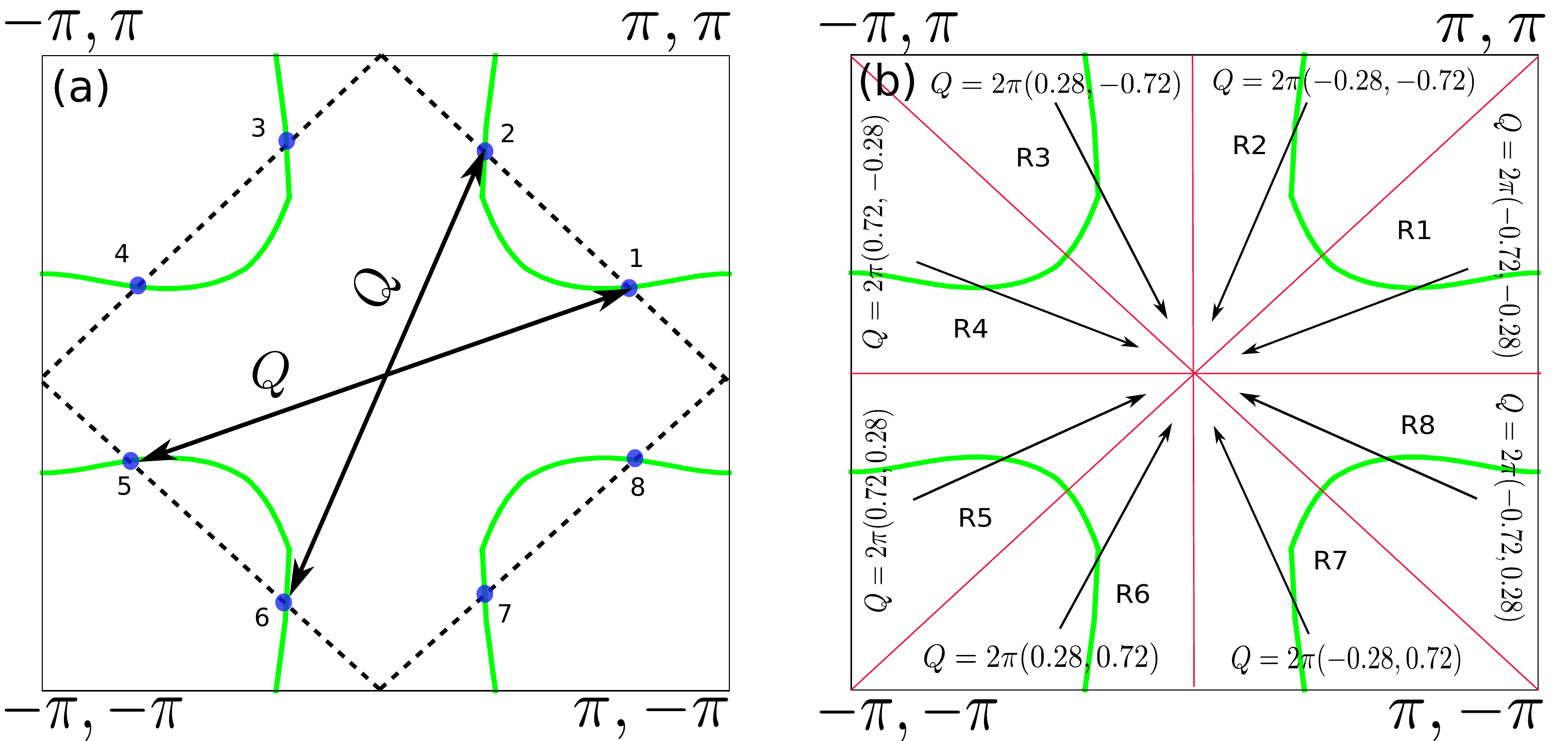} \caption{Diagonal CDW. (a) Green lines show the Fermi surface of the underdoped YBCO band in the first Brillouin zone. The antiferromagnetic Brillouin zone is shown by black dashed lines. `Hot spots', defined by $k$-points where the antiferromagnetic Brillouin zone intersects the Fermi surface, are marked as blue dots and numbered 1-8. CDW wave vectors $Q$ are indicated by arrows, connecting diagonally opposite `hot spots'. (b) The Brillouin zone is divided into eight regions marked R1-R8 to preserve $C_4$ lattice rotational symmetry. All $k$-points in a particular region have same diagonal $Q$ vector, with directions and values indicated in each region.}
\label{fig:FS_diagonalCDW} 
\end{figure}
The `hot spots' theory \cite{Metlitski10b,Efetov13} was originally constructed with wave vectors connecting diagonal parts of the Fermi surface. As shown in Fig.~\ref{fig:FS_diagonalCDW}(a), the $Q$ vectors are here given by the ones connecting `hot spots' marked by `2' (`1') and `6' (`5'), lying on diagonally opposite parts of the Fermi surface. However, the currently experimentally observed magnitude of the CDW wave vectors do not match with the ones proposed in the `hot spots' model. Instead, experiments observe the CDW $Q$ to be along the axial directions \cite{Frano20,Comin16}. Still, as a first example, in this section, we start with the original diagonal CDW $Q$ vector and consider the possibility of the OPDW, before continuing the discussion of the currently experimentally relevant CDW wave vectors in Secs.~\ref{sec:bi-axial}-\ref{sec:robusttobands}. The CDW order in the `hot spots' theory does not break the $C_4$ lattice rotational symmetry. Thus, to ensure the $C_4$ symmetry, we divide the Brillouin zone into 8 regions marked as `R1' to `R8' in Fig.~\ref{fig:FS_diagonalCDW}(b). Each of these octants has a $Q$ connecting different `hot spots'. Directions and values of $Q$ vectors are shown in Fig.~\ref{fig:FS_diagonalCDW}(b).

\begin{figure}[htb]
\includegraphics[width=0.8\linewidth]{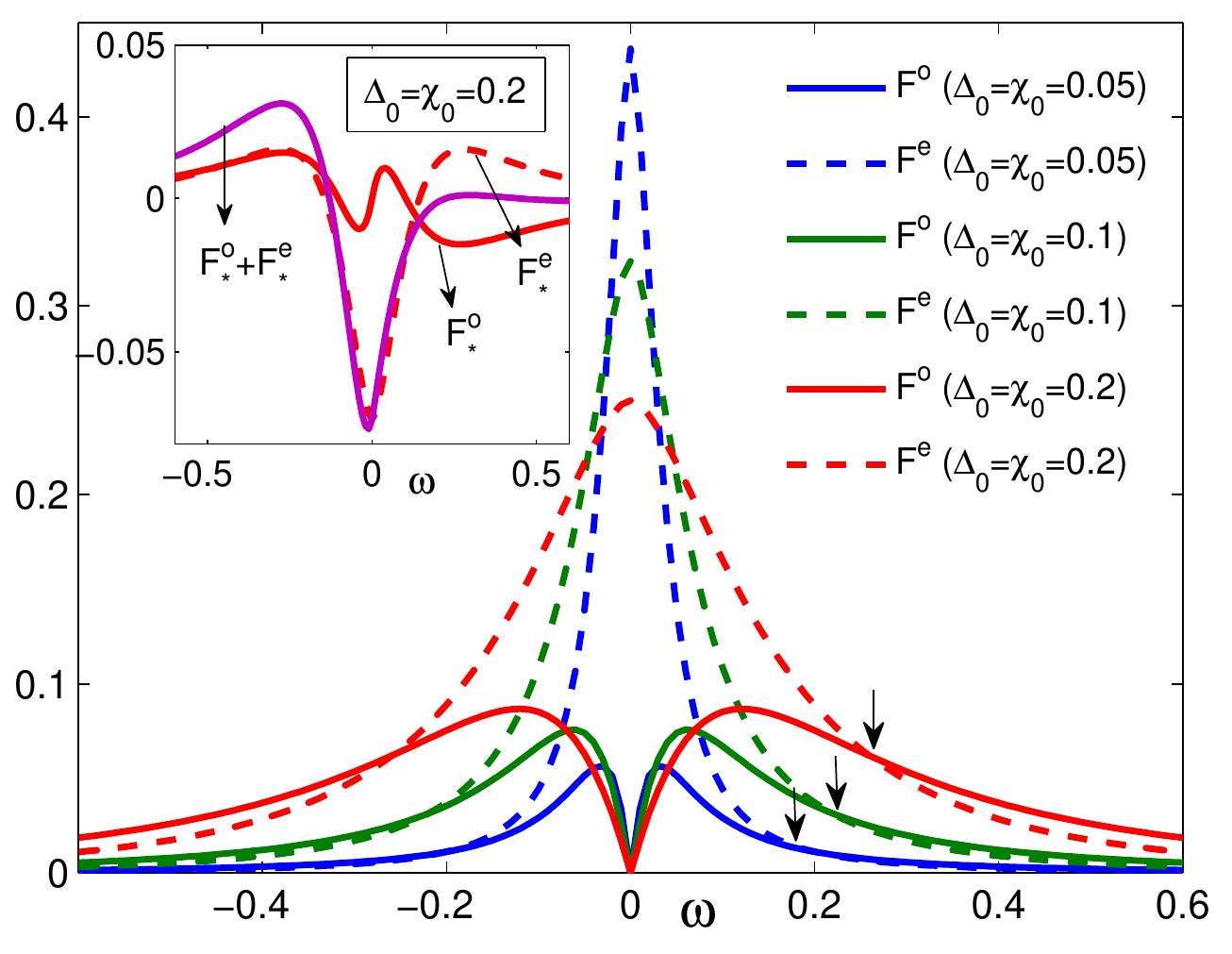} 
\caption{Momentum-averaged absolute values of EPDW, $F^{e}$, (dashed lines) and OPDW, $F^{o}$, (solid lines) plotted as a function of frequency $\omega$ with diagonal $Q$ CDW given in Fig.~\ref{fig:FS_diagonalCDW} for different realistic values of $\Delta_0=\chi_0$.  Arrows indicate the $\omega$ values above which $F^{o}>F^{e}$. Inset: EPDW (red dashed line), OPDW (red solid line) and the total PDW (magenta solid line) obtained when the momentum averaging is done considering the actual sign. The odd-frequency behavior of the OPDW is apparent in the inset. Due to the odd-frequency behavior of the OPDW, the total PDW has a clear asymmetry between positive and negative $\omega$. $\omega$ is in units of $t_1$.}
\label{fig:res_diagonalCDW} 
\end{figure}
Using the diagonal CDW wave vectors in Fig.~\ref{fig:FS_diagonalCDW}, we plot in Fig.~\ref{fig:res_diagonalCDW}, the PDW amplitudes $F^{e}$ and $F^{o}$ for different realistic values \cite{Schabel98,Kondo06,Vishik12,Norman07,Torre16} of $\Delta_0=\chi_0$. We only show results taking $\Delta_0=\chi_0$ for two reasons. First, recent Raman measurements \cite{Loret19,Loret20} suggest that the energy scales corresponding to the superconducting and CDW order parameters are very close to each other for a large range of doping levels for different cuprate materials. Second, from Eq.~\eqref{eq:selfcon}, the relative values of $F_{k}^{o}(i\omega)$ and $F_{k}^{e}(i\omega)$ do not depend on $\Delta_0$ or $\chi_0$, for a fixed $k$ and $\omega$. 
We see in Fig.~\ref{fig:res_diagonalCDW} that $F^{e}$ acquires its maximum value for $\omega=0$ and we call this value $F^{e}_{\text{max}}$. On the other hand, $F^{o}$ is zero at $\omega=0$ by definition of being odd in $\omega$. $F^{o}$ instead peaks for low but finite $\omega$ and we call this value $F^{o}_{\text{max}}$. Even for the small value of $\Delta_0=0.05$, we note that $F^{o}$ is finite, although $F^{o}_{\text{max}}$ is small compared to $F^{e}_{\text{max}}$. However, values of $F^{o}$ and $F^{e}$ become very similar for $\omega$ greater than a particular value indicated by vertical arrows in Fig.~\ref{fig:res_diagonalCDW}. Notably, we also find that increasing $\Delta_0$ increases $F^{o}_{\text{max}}$, while the same decreases $F^{e}_{\text{max}}$. Therefore, the ratio of $F^{o}_{\text{max}}$ to $F^{e}_{\text{max}}$ is strongly increased for stronger SC. For example, for $\Delta_0=0.2$, $F^{o}_{\text{max}}$ is almost half of $F^{e}_{\text{max}}$.  In the inset of Fig.~\ref{fig:res_diagonalCDW}, we also show $F^{o}_{*}$, $F^{e}_{*}$, and $F^{o}_{*}+F^{e}_{*}$, which are the total momentum sums of the OPDW, the EPDW, and the total PDW, respectively, when considering their signs, as defined in Eq.~\eqref{eq:EOPDW_ksum}. The large differences in the values of $F^{o/e}_{*}$ and $F^{o/e}$ already suggests that there are notable sign changes in both PDWs. The total PDW shows a considerable asymmetry in $\omega$ due to the contribution from the OPDW.

\begin{figure}[htb]
\includegraphics[width=1.0\linewidth]{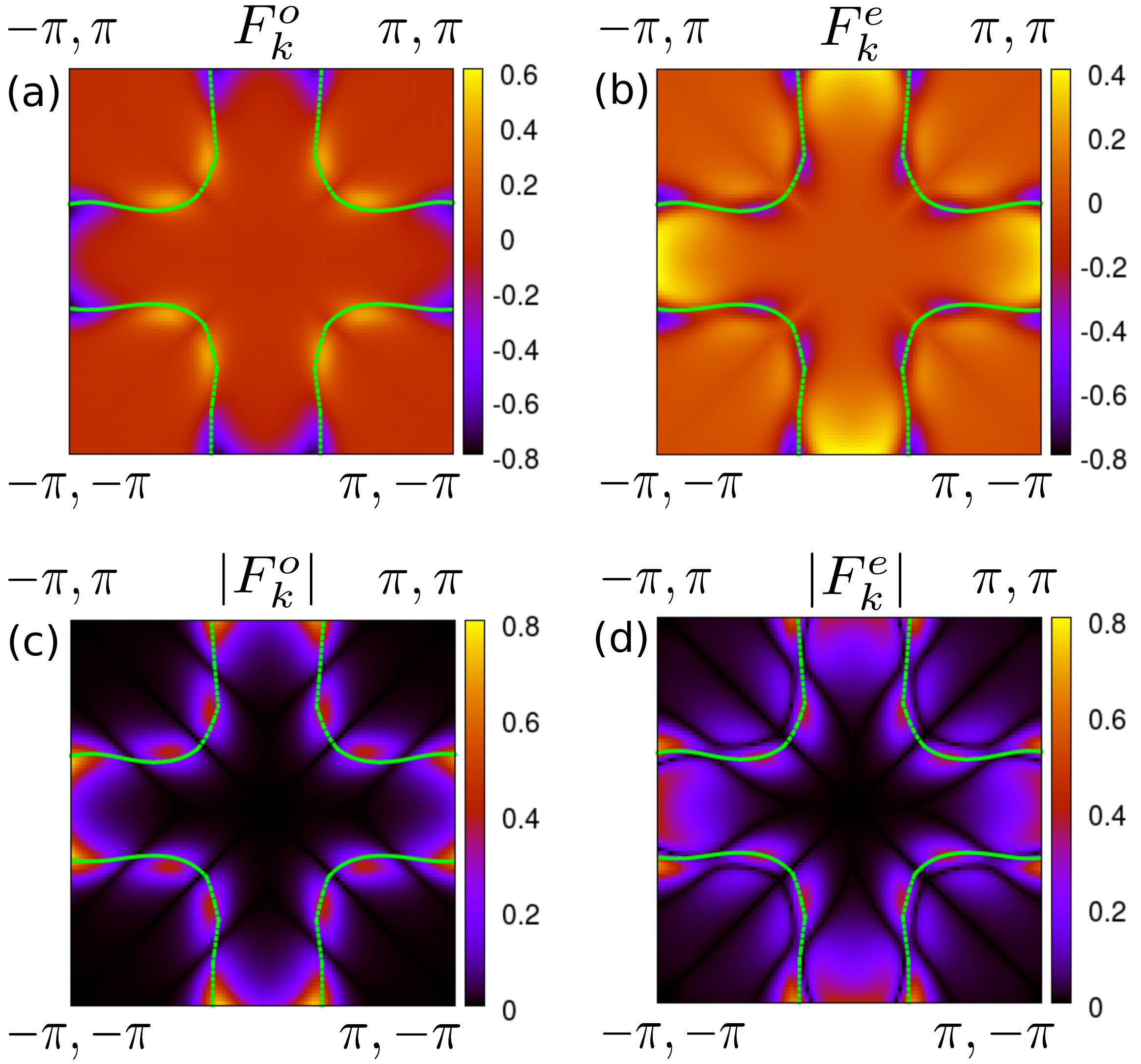} \caption{Color density map of the momentum structure of (a) OPDW, $F^{o}_k$, and (b) EPDW, $F^{e}_k$, with $\Delta_0=\chi_0=0.2$ at $\omega=0.19$ with diagonal $Q$ CDW given in Fig.~\ref{fig:FS_diagonalCDW}. The absolute values (c) $\left|F^{o}_k\right|$ and (d) $\left|F^{e}_k\right|$ give a clearer visualization of the momentum structure for the same parameters as in (a) and (b). The Fermi surface is overlaid with  green lines.}
\label{fig:res_diagonalmomentum} 
\end{figure}
To understand the sign change and the detailed momentum structure of the induced PDW, we plot color density maps of $F_{k}^{o}(i\omega)$ and $F_{k}^{e}(i\omega)$ in Fig.~\ref{fig:res_diagonalmomentum} for $\Delta_0=\chi_0=0.2$ at a specific frequency, $\omega=0.19$. We choose these parameters because the momentum averaged OPDW and EPDW in Fig.~\ref{fig:res_diagonalCDW} are of comparable magnitude, and the qualitative features of $F_{k}^{o}(i\omega)$ and $F_{k}^{e}(i\omega)$ do not change with $\omega$.
We also overlay the Fermi surface of the YBCO band considered in green. We see that the OPDW $F_{k}^{o}$ in Fig.~\ref{fig:res_diagonalmomentum}(a) changes sign along the Fermi surface, going from the anti-nodal region (near ($\pi,0$) and three other $C_4$ symmetric regions) to the nodal region (near $k_x=k_y$ or $k_x=-k_y$ lines). In contrast, $F_{k}^{e}$ in Fig.~\ref{fig:res_diagonalmomentum}(b) does not change signs along the Fermi surface, instead it changes sign as we go away from the Fermi surface. Plotting the absolute values, $\left|F_{k}^{o}\right|$ and $\left|F_{k}^{e}\right|$ in Figs.~\ref{fig:res_diagonalmomentum}(c) and (d), respectively, help to better visualize the zeros. Since $\chi_k$ is assumed to have $d$-wave character, both $F_{k}^{e}$ and $F_{k}^{o}$ have zeros along the $k_x=k_y$ or $k_x=-k_y$ lines. Additionally, $F_{k}^{o}$ is also zero at `hot spots', since $\Delta_{k+Q}=\Delta_k$, giving sign change across these spots. Despite the difference in nodal structure, the regions of the Brillouin zone with high values of PDW correlations are very similar for both the even- and odd-frequency parts.

\subsection{Bi-axial CDW}\label{sec:bi-axial}

\begin{figure}[htb]
\includegraphics[width=1.0\linewidth]{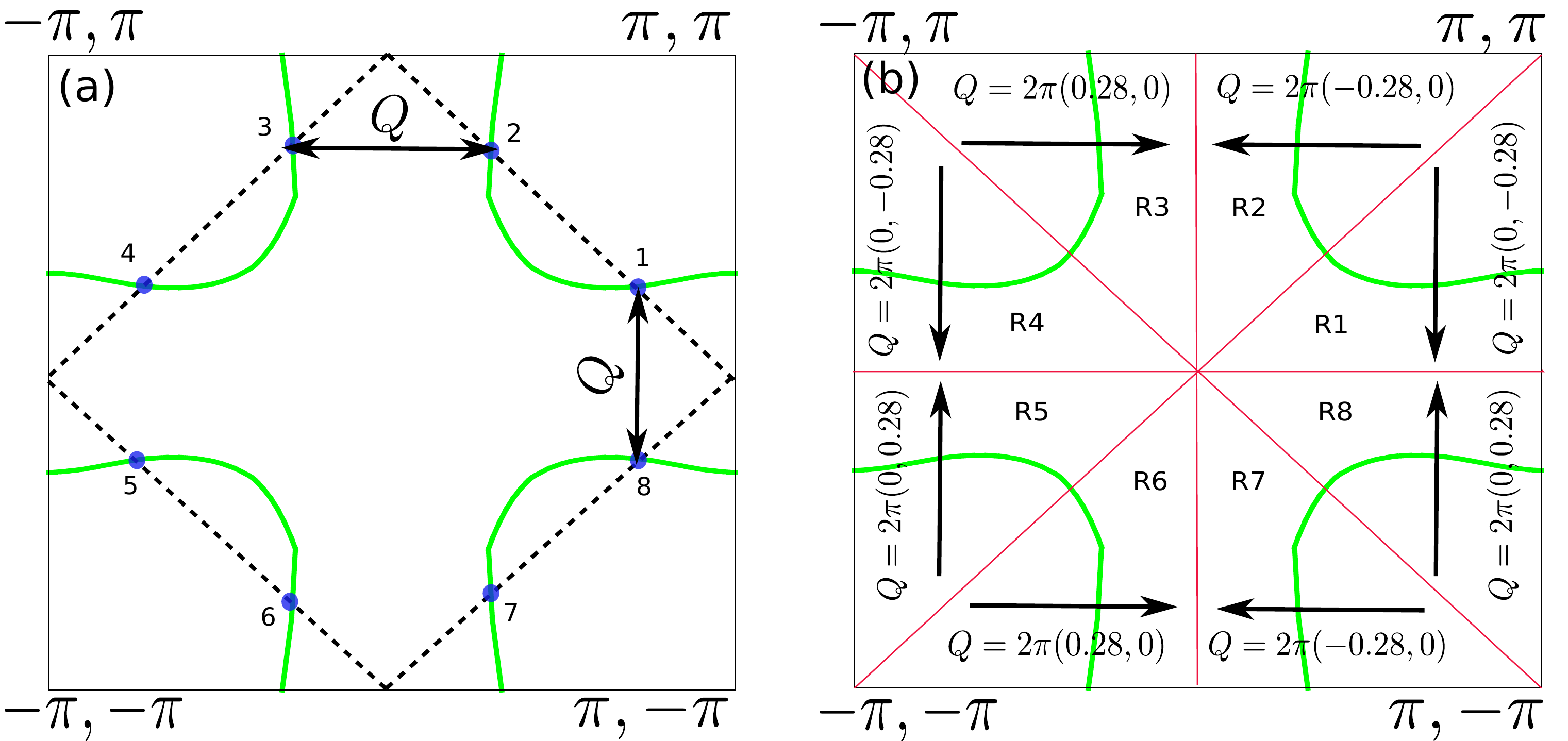} \caption{Bi-axial CDW. (a) Green lines show the same Fermi surface as in Fig.~\ref{fig:FS_diagonalCDW} with `hot spots' denoted as blue dots. CDW wave vectors $Q$ are indicated by arrows,  taken to be parallel to the crystallographic axis connecting nearest `hot spots' with $Q$ vector along $x$-axis for `hot spots' 2,3,6,7 and along $y$-axis for `hot spots' 4,5,8,1. (b) The Brillouin zone is divided into the same eight regions as in Fig.~\ref{fig:FS_diagonalCDW}, but now with axial $Q$ vectors, with direction and magnitude indicated in each region.}
\label{fig:FS_bi-axialCDW} 
\end{figure}
Although a diagonal CDW is the primary charge instability in models with short-range antiferromagnetic interactions \cite{Metlitski10b,Efetov13}, recent experiments in cuprates suggest that the CDW are actually axial in nature \cite{Frano20,Comin16}. In particular, at low or zero magnetic fields, the CDW $Q$ in YBCO is found to be bi-axial with $Q=(Q_x,0)$ and $Q=(0,Q_y)$ \cite{daSilvaNeto14,Comin14}. The magnitudes of the observed $Q$ vectors are very close to the wave vectors connecting neighboring `hot spots' \cite{Comin14}, as shown in Fig.~\ref{fig:FS_bi-axialCDW}(a). CDW with such axial wave vectors have also been found as a competing instability in models with antiferromagnetic interactions \cite{Wang14} and can furthermore be enhanced by including an off-site Coulomb interaction \cite{Allais14c,Chakraborty19}. Thus, in this section, we consider the CDW wave vector to be bi-axial connecting neighboring `hot spots' as in Fig.~\ref{fig:FS_bi-axialCDW}(a). As there is no experimental evidence that bi-axial CDW breaks the $C_4$ lattice rotational symmetry, we again separate the Brillouin zone into eight regions as shown in Fig.~\ref{fig:FS_bi-axialCDW}(b) to ensure the $C_4$ symmetry. It should also be noted that the lattice structure of YBCO is strongly orthorhombic, thus breaking $C_4$ symmetry \cite{shaked94}. However, since we here consider a particular electronic instability with no evidence of breaking $C_4$ symmetry, we ignore the $C_4$ symmetry breaking in the lattice structure of YBCO.

\begin{figure}[htb]
\includegraphics[width=0.8\linewidth]{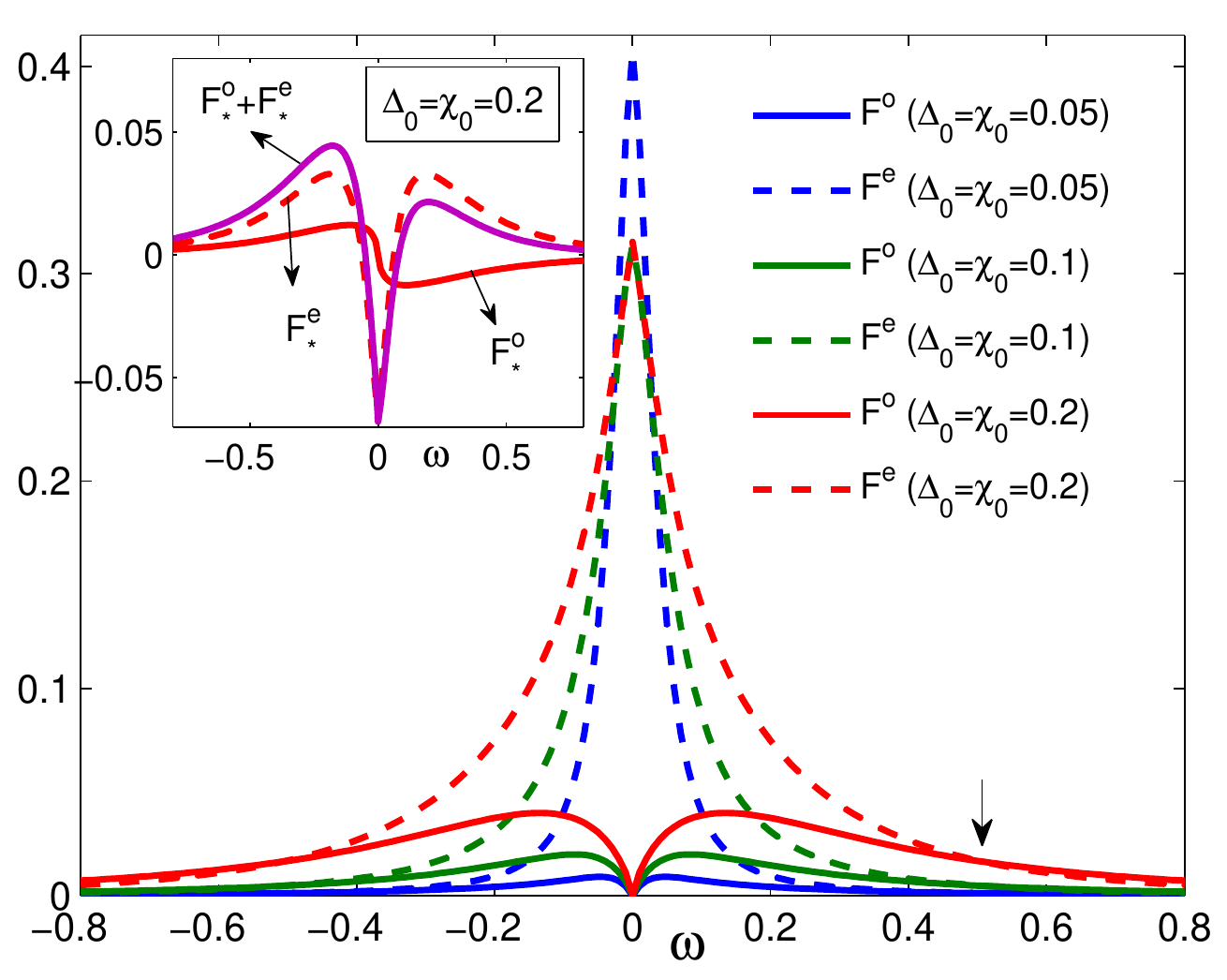} \caption{Momentum-averaged absolute values of EPDW, $F^e$, (dashed lines) and OPDW, $F^o$, (solid lines) plotted as a function of frequency $\omega$ with bi-axial $Q$ CDW given in Fig.~\ref{fig:FS_bi-axialCDW}, and for the same realistic values of $\Delta_0 =\chi_0$ as in Fig.~\ref{fig:res_diagonalCDW}. Arrow indicates the $\omega$ value above which $F^{o}>F^{e}$. Inset: EPDW (red dashed line), OPDW (red solid line) and the total PDW (magenta solid line) obtained when the momentum averaging is done considering the actual sign. $\omega$ is in units of $t_1$.}
\label{fig:res_bi-axialCDW} 
\end{figure}
We plot the frequency dependence of the EPDW $F^{e}$ and the OPDW $F^{o}$ with bi-axial CDW wave vectors in Fig.~\ref{fig:res_bi-axialCDW}, for the same values of $\Delta_0=\chi_0$ as used for the diagonal CDW. We again find that for all $\Delta_0$, $F^{o}$ attains a finite value, with a frequency dependence that is very similar to the case of diagonal CDW. With increasing $\Delta_0$, the $F^{o}_{\text{max}}$ increases. The $F^{e}_{\text{max}}$, on the other hand, initially decreases with an increase in $\Delta_0$, but does not change with further increase in $\Delta_0$. The ratio of the $F^{o}_{\text{max}}$ to the $F^{e}_{\text{max}}$ therefore increases with increasing $\Delta_0$, but the ratio is, however, somewhat smaller compared to the case with diagonal CDW case reported in Fig.~\ref{fig:res_diagonalCDW}. Still, values of $F^{e}_{*}$ and $F^{o}_{*}$, shown in the inset of Fig.~\ref{fig:res_bi-axialCDW}, are comparable and also illustrate the very different frequency dependencies of the EPDW and OPDW states. This difference in the frequency dependencies gives rise to an asymmetric frequency dependence of $F^{o}_{*}+F^{e}_{*}$, as shown in the inset of Fig.~\ref{fig:res_bi-axialCDW}.

\begin{figure*}[ht]
\includegraphics[width=0.8\linewidth]{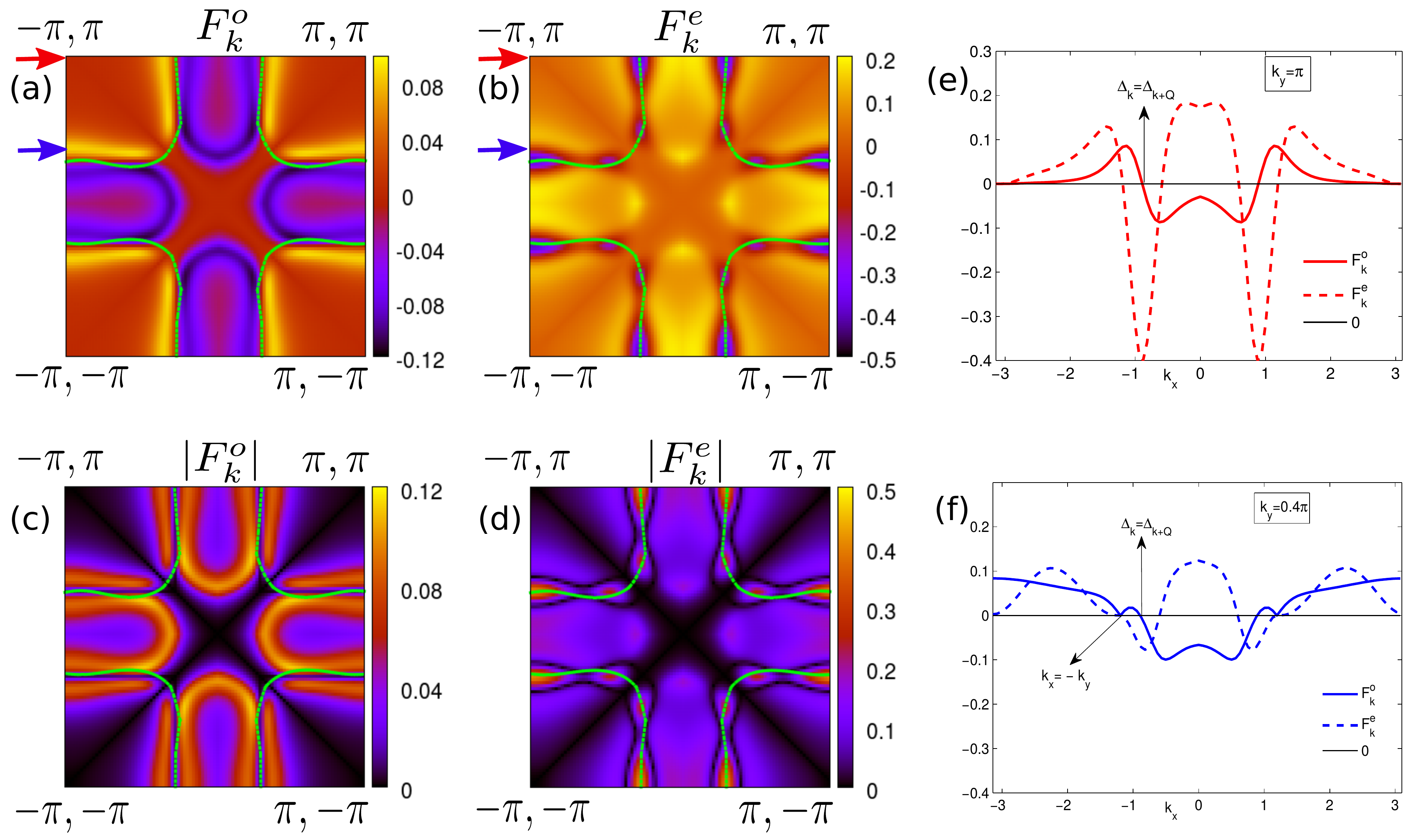} \caption{Color density map of the momentum structure of (a) OPDW, $F^o_k$, and (b) EPDW, $F^e_k$, with $\Delta_0=\chi_0=0.2$ and $\omega=0.19$ with bi-axial $Q$ CDW given in Fig.~\ref{fig:FS_bi-axialCDW}, with absolute values given in (c) and (d), and Fermi surface overlaid with green lines, all similar to Fig.~\ref{fig:res_diagonalmomentum}.
(e) Line cut of EPDW (dashed line) from (a) and OPDW (solid line) from (b) along the line $k_y=\pi$, indicated by red arrows in (a) and (b). (f) Line cut of EPDW (dashed line) from (a) and OPDW (solid line) from (b) along the line $k_y=0.4\pi$, indicated by blue arrows in (a) and (b). }
\label{fig:res_bi-axialmomentum} 
\end{figure*}
We again gain additional insights into the relative strengths of $F_{k}^{o}$ and $F_{k}^{e}$ by looking at their momentum structure. Color density maps of $F_{k}^{o}$, $F_{k}^{e}$, $\left|F_{k}^{o}\right|$ and $\left|F_{k}^{e}\right|$ for $\omega=0.19$ and $\Delta_0=0.2$ are plotted in Fig.~\ref{fig:res_bi-axialmomentum} (a), (b), (c), and (d), respectively. Since the choice of $Q$ is different here compared to the diagonal CDW, lines where $\Delta_{k+Q}=\Delta_k$ also change. This gives a very different momentum space structure, especially for $F_{k}^{o}$. For example, in the region `R2', $\Delta_{k+Q}=\Delta_k$ along a line $k_x=k_x^{\text{HS}}$ parallel to the k$_y$-axis, where $k_x^{\text{HS}}$ is the $x$-coordinate of `hot spot' 2, see Fig.~\ref{fig:FS_bi-axialCDW}. In contrast, the line $\Delta_{k+Q}=\Delta_k$ in the diagonal CDW case is not parallel to the k$_y$-axis. Similar features are observed in the other $C_4$ symmetric regions of the Brillouin zone. $F_{k}^{e}$ shows a much more similar momentum space structure to the diagonal CDW case. In addition, both $F_{k}^{o}$ and $F_{k}^{e}$ have usual zeros along nodal directions, i.e.~along $k_x=\pm k_y$.

To provide a better understanding of the relative magnitude of the EPDW and OPDW in different parts of the Brillouin zone, we make two line cuts along the red and blue arrows in Fig.~\ref{fig:res_bi-axialmomentum}(a,b). $F_{k}^{o}$ and $F_{k}^{e}$ are shown along the red arrow ($k_y=\pi$) in Fig.~\ref{fig:res_bi-axialmomentum}(e) and along the blue arrow ($k_y=0.4\pi$) in Fig.~\ref{fig:res_bi-axialmomentum}(f). In the anti-nodal region, $k_y=\pi$, the induced PDW is clearly dominated by the even-frequency component with the maximum of OPDW being at most 25\% of the EPDW. However, close to the nodal region ($k_y=0.4\pi$), the OPDW and EPDW have very similar maximum values. 
Thus, even if momentum integrated values of $F_{k}^{o}$ are small compared to their even-frequency counterparts, as indicated in Fig.~\ref{fig:res_bi-axialCDW}, the magnitudes of OPDW and EPDW become clearly very comparable in some parts of the Brillouin zone. For example, at $k_y=0.4\pi$, the maximum magnitude of the OPDW is obtained for $k_x=0.16\pi$. At this $k$ point, $\left|F^{o}_{k}\right| \approx 0.1$, $\left|F^{e}_{k}\right| \approx 0.05$, and $\left|\Delta_{k}\right|=\left|\chi_{k}\right| =0.1$. Thus, not only is the OPDW twice the magnitude of the EPDW, it is also of equal magnitude of the uniform SC or CDW from which it is derived.

\subsection{Uni-axial CDW}\label{sec:uniaxial}

\begin{figure}[htb]
\includegraphics[width=1.0\linewidth]{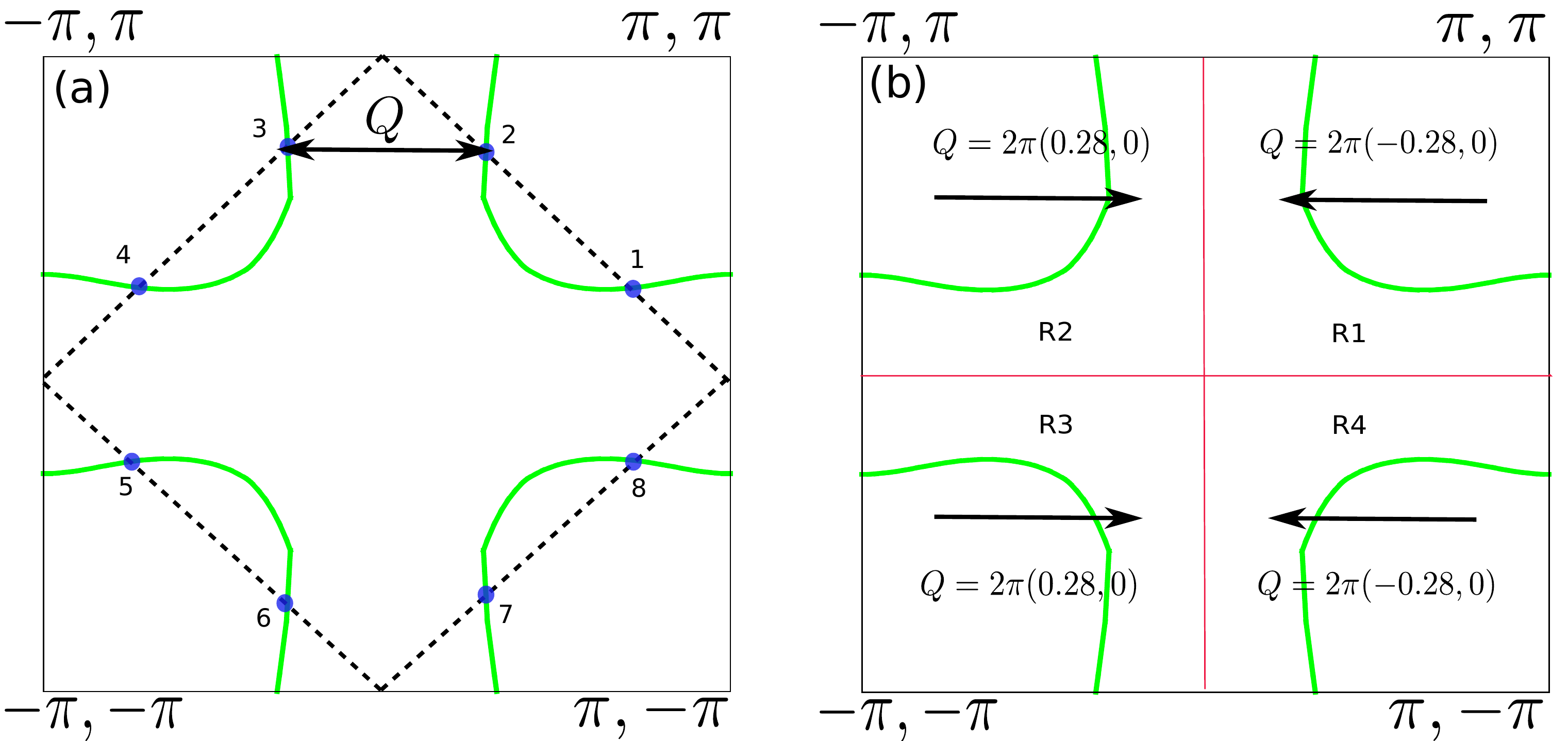} \caption{Uni-axial CDW. (a) Green lines show the same Fermi surface as in Fig.~\ref{fig:FS_diagonalCDW} with `hot spots' denoted as blue dots. CDW wave vector $Q$ is indicated by an arrow, taken to be only along the $x$-axis, hence called uni-axial, but with same magnitude as in Fig.~\ref{fig:FS_bi-axialCDW}. (b) The Brillouin zone is divided into four regions marked R1-R4 to ensure  broken $C_4$ symmetry but preserved $C_2$ symmetry. All $k$-points in a particular region have same diagonal $Q$ vector, with directions and values indicated in each region.}
\label{fig:FS_uniaxialCDW}
\end{figure}
The bi-axial CDW discussed in the previous section is observed in YBCO, only at zero or low magnetic fields \cite{Ghiringhelli12,daSilvaNeto14,Comin14}. A change in the nature of CDW, however, occurs as one applies a strong magnetic field $B$. For $B>17$~T, the correlation length of CDW jumps from $\sim 20$ to $\sim 100$ lattice spacings, indicating a transition to a `true' long-range CDW phase \cite{Gerber:2015gx,Chang16,LeBoeuf13}. Long-range CDW are uni-axial in nature with $Q=(Q_x,0)$ or $Q=(0,Q_y)$, but not both. In other words, a transition occurs from a checkerboard CDW to a stripe CDW with increasing magnetic field. Long-range CDW have also been found to coexist with SC in a window of magnetic field $B_{c2}>B>17$~T \cite{LeBoeuf13,Grissonnanche14,Zhou17}, where $B_{c2}$ is the upper critical field of the superconducting order. Although the magnetic field introduces vortex-induced inhomogeneities in a strongly type-II cuprate superconductor, an effective homogeneous Hamiltonian, as in Eq.~\eqref{eq:Hamil}, still gives a reasonable description of the coexistence phase close to $B_{c2}$ \cite{Tinkhambook, DeGennesbook, Rosenstein10, Chakraborty18}. We therefore also study uni-axial CDW, where we consider $Q$ vectors only along one axis as shown in Fig.~\ref{fig:FS_uniaxialCDW}(a). The uni-axial nature of CDW breaks the $C_4$ rotational symmetry of the Brillouin zone. We thus here need to separate the Brillouin zone into four regions, as shown in Fig.~\ref{fig:FS_uniaxialCDW}(b), instead of eight for the previously treated CDW.

\begin{figure}[htb]
\includegraphics[width=0.8\linewidth]{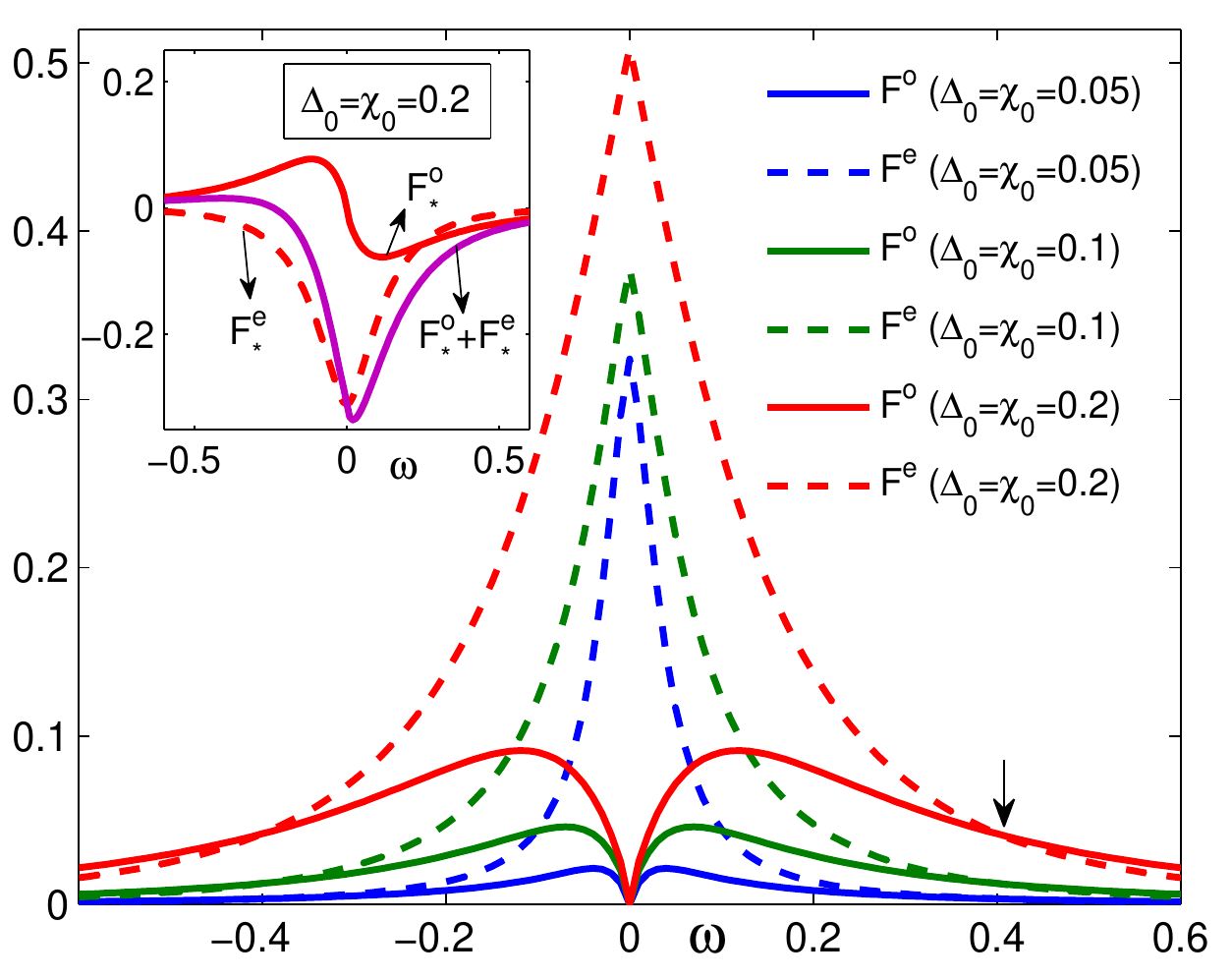} \caption{Momentum-averaged absolute values of EPDW, $F^e$, (dashed lines) and OPDW, $F^o$ (solid lines) plotted as a function of frequency $\omega$ with uni-axial $Q$ CDW in Fig.~\ref{fig:FS_uniaxialCDW} and for the same realistic values of $\Delta_0 =\chi_0$ as in Fig.~\ref{fig:res_diagonalCDW}. Arrow indicates the $\omega$ value above which $F^o>F^e$. Inset: EPDW (red dashed line), OPDW (red solid line) and the total PDW (magenta solid line) obtained when the momentum averaging is done considering the actual sign. $\omega$ is in units of $t_1$.}
\label{fig:res_uniaxialCDW} 
\end{figure}
Similar to the findings in the two previous sections for diagonal and bi-axial CDW, the OPDW $F^{o}$ increases with increasing $\Delta_0$ also for a uni-axial CDW, as shown in Fig.~\ref{fig:res_uniaxialCDW}. Interestingly, $F^{o}$ with uni-axial CDW wave vectors show a considerable increase compared to $F^{o}$ with bi-axial CDW (cp.~Fig.~\ref{fig:res_bi-axialCDW}) for all values of $\Delta_0$, keeping $\Delta_0$ and $\chi_0$ unchanged in the two cases. For example, for $\Delta_0=0.2$, $F^{o}_{\text{max}}\approx 0.1$ in Fig.~\ref{fig:res_uniaxialCDW}, whereas $F^{o}_{\text{max}}\approx 0.04$ for a bi-axial CDW in Fig.~\ref{fig:res_bi-axialCDW}. Compared to the other investigated CDW, $F^{e}_{\text{max}}$ for a uni-axial CDW behaves differently and increases with increasing $\Delta_0$. Thus, even though the $F^{o}_{\text{max}}$ is increased with a uni-axial CDW, the ratio of the $F^{o}_{\text{max}}$ to the $F^{e}_{\text{max}}$ remains about same as for the bi-axial CDW. Notably, for high $\omega$ and $\Delta_0$, the OPDW becomes comparable and even larger than the EPDW. For example, $\Delta_0=0.2$ gives $F^{o}>F^{e}$ for all $\omega>0.4$ and $F^{o}_{*}>F^{e}_{*}$ for $\omega>0.25$.

\begin{figure}[htb]
\includegraphics[width=1.0\linewidth]{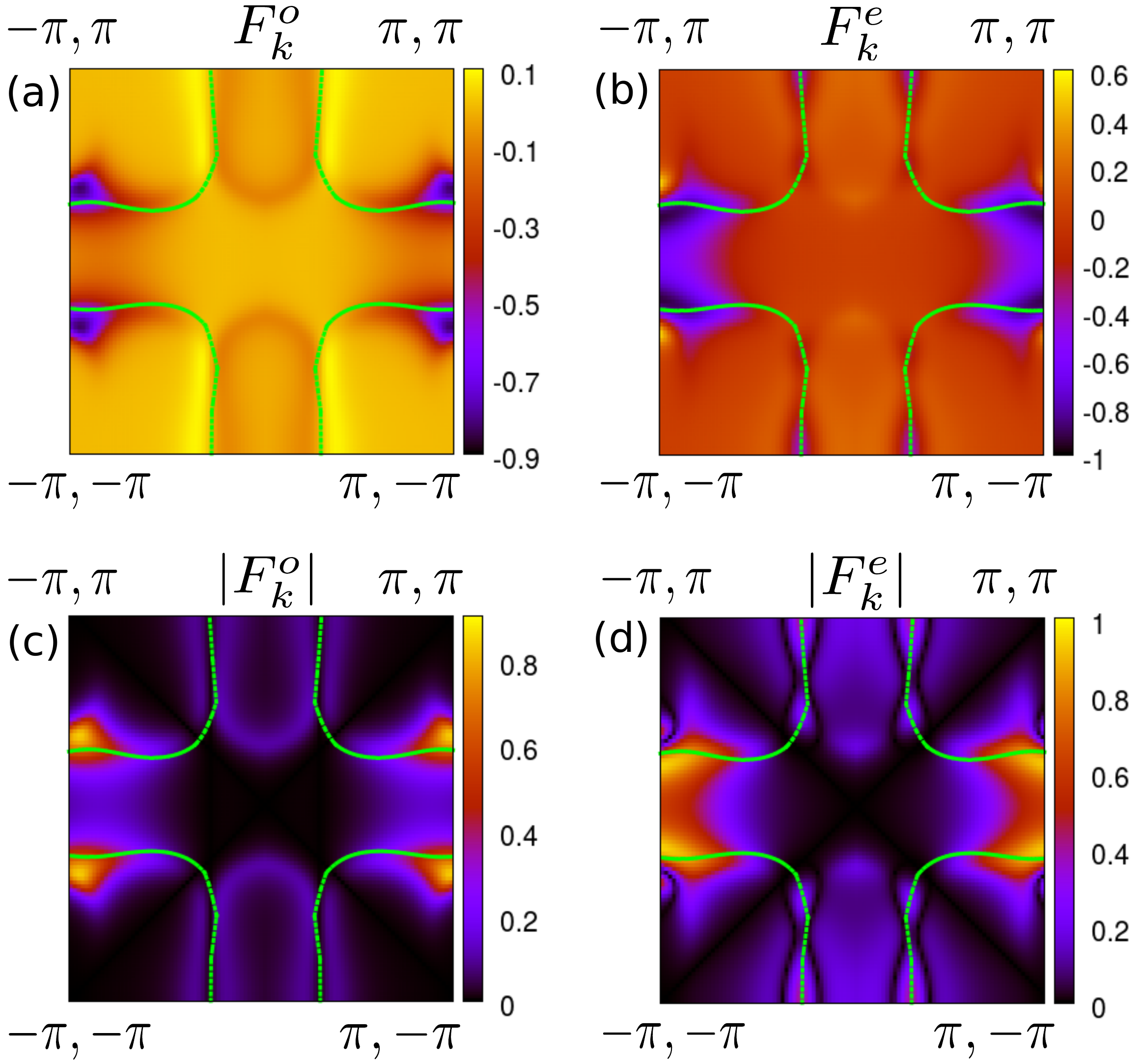} \caption{Color density map of the momentum structure of (a) OPDW, $F^o_k$, and (b) EPDW, $F^e_k$, with $\Delta_0=\chi_0=0.2$ and $\omega=0.19$ with uni-axial $Q$ CDW given in Fig.~\ref{fig:FS_uniaxialCDW}, with absolute values given in (c) and (d), and Fermi surface overlaid with green lines, all similar to Fig.~\ref{fig:res_diagonalmomentum}. Note that the maximum values of the color bars in (c) and (d) are much higher compared to Figs.~\ref{fig:res_bi-axialmomentum}(c,d).}
\label{fig:res_uniaxialmomentum} 
\end{figure}
The broken $C_4$ symmetry with a uni-axial CDW is clearly reflected in the momentum structure of the OPDW and the EPDW plotted Fig.~\ref{fig:res_uniaxialmomentum}. In the anti-nodal regions near $k_y=\pm \pi$, both $F_{k}^{o}$ and $F_{k}^{e}$ behave similar to the case of bi-axial CDW, as the choice of $Q$ is the same in these regions for the uni-axial and bi-axial CDW. The distinction between uni-axial CDW and bi-axial CDW instead comes in the anti-nodal regions near $k_x=\pm \pi$. Values of $F_{k}^{o}$ and $F_{k}^{e}$ near $k_x=\pm \pi$ are significantly enhanced compared to the values near $k_y=\pm \pi$, with both displaying very high values near the Fermi surface. We note that $\left|F_{k}^{o}\right|$ decays faster than $\left|F_{k}^{e}\right|$ as we go away from the Fermi surface to a region close to $(\pm \pi,0)$, giving $\left|F_{k}^{o}\right|<\left|F_{k}^{e}\right|$ at $(\pm \pi,0)$. Thus, after momentum averaging, $F^{o}<F^{e}$ as found in Fig.~\ref{fig:res_uniaxialCDW}, but where the two different PDW show very similar values around the Fermi surface.

\subsection{Band structure robustness}\label{sec:robusttobands}
\begin{figure}[htb]
\includegraphics[width=1.0\linewidth]{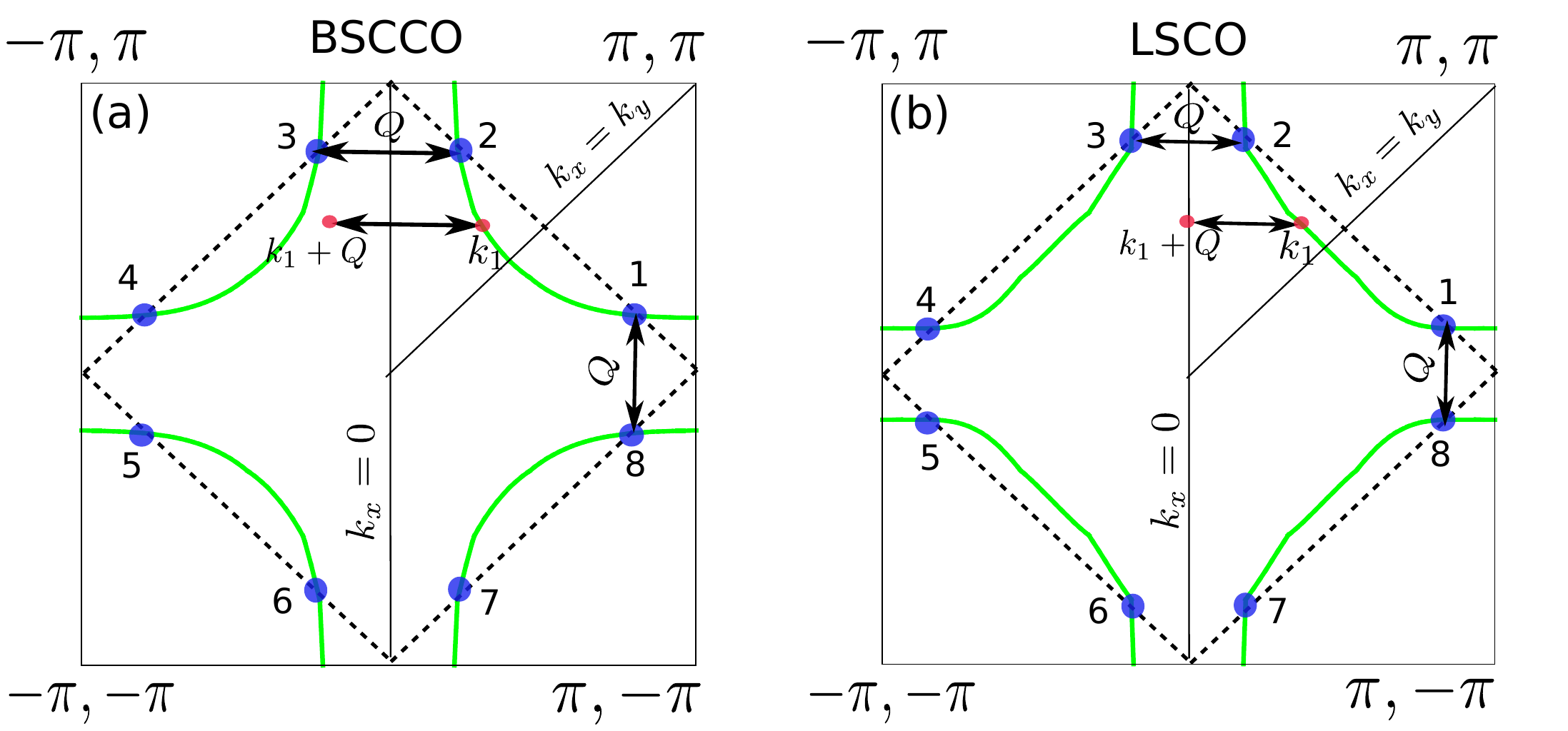} \caption{Green lines show the Fermi surface of the BSCCO band (a) and the LSCO band (b) in the first Brillouin zone. `Hot spots' are marked as blue dots and Bi-axial CDW wave vectors $Q$ are indicated by arrows. Note that the magnitudes of $Q$ are determined by the distance between neighboring `hot spots' and are different from the value in each material. A representative $k$-point, $k_1$, on the Fermi surface is marked by a red dot, with $k_1+Q$ also marked with a red point.}
\label{fig:bands} 
\end{figure}

Our results above for different choices of $Q$ vectors for the CDW show that the induced OPDW is a common feature when there exists a coexistence of SC and CDW. One might, however, ask how sensitive these results are to the band structure considered. Till now, we have considered only a YBCO band structure, see Eq.~\eqref{eq:Band1parameters}. In this section we further investigate two different bands with contrasting features. In this endeavor, we take the parameters which mimic the band structures of underdoped Bi$_2$Sr$_2$CaCu$_2$O$_{8+x}$ (BSCCO) and La$_{2-x}$Sr$_x$CuO$_4$ (LSCO) \cite{Norman07}. All the three band parameters are given in Table.~\ref{tab:table1}. 
\begin{table}[t]
 \begin{tabular}{|c | c | c| c|}
 \hline
  Band Parameters  & YBCO & BSCCO & LSCO  \\
 \hline
 $t_1$ & -1.1259 & -0.6798 & -0.7823 \\
 $t_2$ & 0.5540 & 0.2368 & 0.0740 \\
$t_3$ & -0.1174 & -0.0794 & -0.0587 \\
 $t_4$ & -0.0701 & 0.0343 & -0.1398 \\
 $t_5$ & 0.1286 & 0.0011 & -0.0174 \\
 $\mu$ & 0.1756 & 0.196 & 0.0801 \\
 \hline
\end{tabular}
\caption{Band parameters of the dispersion given in Eq.~\eqref{eq:Band} corresponding to three different cuprate materials. All the parameters are in units of eV. \label{tab:table1}}
\end{table}
The Fermi surfaces of the BSCCO and LSCO band structures are shown in Fig.~\ref{fig:bands} (a) and (b), respectively. While the BSCCO band has a Fermi surface with isotropic Fermi velocities and no nesting regions, the LSCO band has `approximate' nesting only in a small region near the anti-nodes. In contrast, the earlier considered YBCO band has long `approximately' nested necks around the `hot spots', as seen in Fig.~\ref{fig:FS_diagonalCDW}(a).

\begin{figure}[htb]
\includegraphics[width=1.0\linewidth]{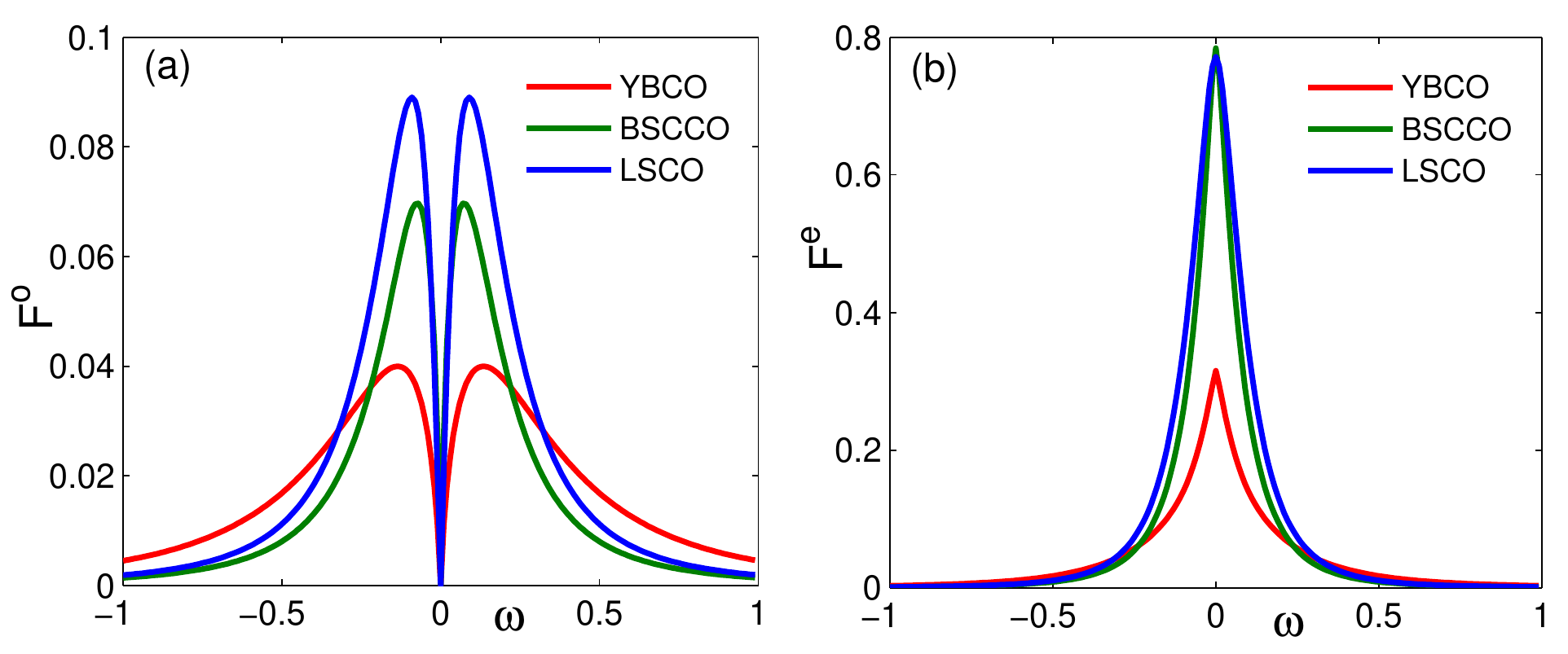} \caption{Momentum-averaged absolute values of (a) OPDW, $F^o$ and (b) EPDW, $F^e$ plotted as a function of frequency $\omega$ with a bi-axial CDW and $\Delta_0=\chi_0=0.2$ for three different band structures representing underdoped YBCO (red), BSCCO (green), and LSCO (blue). $\omega$ is in units of $t_1$.}
\label{fig:res_bandcomparison} 
\end{figure}

As discussed in Secs.~\ref{sec:bi-axial} and \ref{sec:uniaxial}, in YBCO, the long range uni-axial CDW is observed at high magnetic field, whereas the bi-axial CDW is observed at low or zero magnetic fields \cite{Ghiringhelli12,daSilvaNeto14,Comin14}. However, we note that BSCCO shows local uni-axial CDW \cite{Fujita14a,Hamidian15a} and LSCO features uni-axial stripe CDW \cite{Emery99}, even at zero magnetic fields. Still, in order to be able to primarily focus  on the band structure dependencies for the induced PDW, we here consider the case of CDW with bi-axial $Q$ vectors for all three cuprates and thus divide the Brillouin zone into eight regions as in Fig.~\ref{fig:FS_bi-axialCDW}(b). The magnitude of $Q$ is given by the distance between neighboring `hot spots' in the corresponding bands, indicated in Fig.~\ref{fig:bands}. 
We plot the results in Fig.~\ref{fig:res_bandcomparison}, where $F^{o}$ and $F^{e}$ are shown for all three cuprates at $\Delta_0=0.2$. In all three cases, $F^{o}$ is finite and behave very similarly with frequency. $F^{o}_{\text{max}}$ increases notably from YBCO to BSCCO to LSCO. This is primarily due to the fact that $\Delta_{k+Q}-\Delta_{k}$ is larger in most parts of the Fermi surface in BSCCO and LSCO compared to YBCO. In order to illustrate this statement, we focus on a representative point $k_1$ on the Fermi surface, indicated in Fig.~\ref{fig:bands}. For LSCO, $k_1+Q$ is found at $k_x=0$. As a result, $\Delta_{k_1+Q}$ is maximum due to the $d$-wave nature (note that $Q=(-Q_x,0)$ does not change ${k_1}_y$). On the other hand, $k_1+Q$ for BSCCO lies away from $k_x=0$. Although not shown in Fig.~\ref{fig:bands}, $k_1+Q$ for YBCO lies even further away from $k_x=0$. So, we clearly find $\Delta_{k_1+Q} (\text{LSCO})>\Delta_{k_1+Q} (\text{BSCCO})>\Delta_{k_1+Q} (\text{YBCO})$. Furthermore, $\Delta_{k_1} (\text{LSCO})\approx \Delta_{k_1} (\text{BSCCO}) \approx \Delta_{k_1} (\text{YBCO})$ and $F_{k_1}^{o}\propto \Delta_{k_1+Q}-\Delta_{k_1}$. As a direct consequence, $F^{o}_{\text{max}}$ is the largest in LSCO and smallest in YBCO. However, for YBCO, $F_{o}$ decays a bit more slowly with $\omega$, such that its value actually become the highest among the three bands for large frequencies.
We also note that these band structure results show that the OPDW presented in the earlier three sections, \ref{sec:diagonal}-\ref{sec:uniaxial}, is likely underestimated in terms of its magnitude. 
In spite of these band effects, the ratio of $F^{o}_{\text{max}}$ to $F^{e}_{\text{max}}$ does not change with the change in the band structure as $F^{e}_{\text{max}}$ also increases in BSCCO and LSCO. Based on these results for three different band structures, representing three different underdoped cuprates, we conclude that the OPDW is a robust and ubiquitous feature in cuprate superconductors, although there exist some quantitative differences. 

\section{Spin-triplet odd-frequency PDW: Coexistence of SC and SDW}\label{sec:SCSDW}

Having established that CDW generically give rise to OPDW correlations, and especially in the cuprates, we also explore whether a spin density wave (SDW) can induce such correlations. In order to investigate the coexistence of SC and SDW, we write a mean-field Hamiltonian in momentum space,
\begin{eqnarray}
H_{\text{SDW}}&=&\sum_{k,\sigma} \xi_{k} c_{k \sigma}^{\dagger} c_{k \sigma} + \sum_{k, \sigma} \sigma \left( m_{k} c_{k \sigma}^{\dagger} c_{k+Q \sigma} + \textrm{H.c.} \right) \nonumber \\
&+& \sum_{k} \left( \Delta_{k} c_{-k \downarrow} c_{k \uparrow} + \textrm{H.c.} \right)+\text{constant},
\label{eq:Hamil_SDW}
\end{eqnarray}
where $m_k$ is the SDW order parameter with a modulation wave vector $Q$ and all other notation is the same as in Eq.~\eqref{eq:Hamil}. This Hamiltonian can be written in a matrix form, in the basis as for the CDW case: $\Psi^{\dagger}=\left(c_{k \uparrow}^{\dagger},c_{k+Q \uparrow}^{\dagger},c_{-k \downarrow},c_{-k-Q \downarrow}\right)$ as,
\begin{eqnarray}
&&H_{\text{SDW}}=\frac{1}{2}\sum_{k} \Psi^{\dagger} \hat{H}_{\text{SDW}} \Psi+\text{constant}, \nonumber \\
&&\text{with} \nonumber \\
&&\hat{H}_{\text{SDW}}=\left(\begin{array}{cccc} \xi_k & m_{k} & \Delta_{k} & 0 \\
m_k & \xi_{k+Q} & 0 & \Delta_{k+Q} \\
\Delta_k & 0 & -\xi_{k} & m_{k} \\
0 & \Delta_{k+Q} & m_{k} & -\xi_{k+Q} \\
\end{array}\right).
\label{eq:Hamilmat_SDW}
\end{eqnarray} 
We can obtain the Green's function $G^{s}$ for Eq.~\eqref{eq:Hamilmat_SDW} in the same way as in Sec.~\ref{sec:OPDW} and the PDW correlator is given by,
\begin{eqnarray}
&&F^{s}_{k,k+Q}(i\omega)=G^{s}_{14}(i\omega)=F^{s,even}_{k,k+Q}(i\omega)+F^{s,odd}_{k,k+Q}(i\omega),\nonumber \\
&&\text{where}\nonumber \\
&&F^{s,even}_{k,k+Q}(i\omega)=\frac{m_{k}\left(\xi_k\Delta_{k+Q}-\xi_{k+Q}\Delta_k\right)}{D}, \label{eq:feven_SDW}\\
&&F^{s,odd}_{k,k+Q}(i\omega)=\frac{i\omega m_{k}\left(\Delta_k+\Delta_{k+Q}\right)}{D},\label{eq:fodd_SDW} \\
&&D=\left( \omega^2+\xi_{k}^2+\Delta_{k}^2 \right)\left( \omega^2+\xi_{k+Q}^2+\Delta_{k+Q}^2 \right)\nonumber \\
&&-2 m_{k}^2 \left( \xi_k \xi_{k+Q}-\Delta_{k}\Delta_{k+Q} -\omega^2 \right)+m_{k}^4. \label{eq:D_SDW}
\end{eqnarray}
The functional form of $D$ here is same as that in Eq.~\eqref{eq:D}, only with $\chi_k$ replaced by $m_{k}$. Here again, we see that the induced PDW, very generically has both even and odd-frequency components. 
However, we note directly that there is a change of signs in the definitions of the PDW generated by SDW compared to the earlier CDW case; the `$+$' sign in Eq.~\eqref{eq:feven} is now given by a `$-$' sign in Eq.~\eqref{eq:feven_SDW} and the `$-$' sign in Eq.~\eqref{eq:fodd} is now given by a `$+$' sign in Eq.~\eqref{eq:fodd_SDW}. 

To investigate the spin symmetries for the SDW-generated PDW, we look at the $G^{s}_{23}(i\omega)$ component of the Green's function, similarly as in Sec.~\ref{sec:OPDW}. We obtain
\begin{eqnarray}
&&F^{s}_{k+Q,k}(i\omega)=G^{s}_{23}(i\omega)=F^{s,even}_{k+Q,k}(i\omega)+F^{s,odd}_{k+Q,k}(i\omega),\nonumber \\
&&\text{where}\nonumber \\
&&F^{s,even}_{k+Q,k}(i\omega)=\frac{m_{k}\left(-\xi_k\Delta_{k+Q}+\xi_{k+Q}\Delta_k\right)}{D}=-F^{s,even}_{k,k+Q}(i\omega), \nonumber \\
&&\label{eq:fevenm_SDW}\\
&&F^{s,odd}_{k+Q,k}(i\omega)=\frac{i\omega m_{k}\left(\Delta_{k+Q}+\Delta_{k}\right)}{D}=F^{s,odd}_{k,k+Q}(i\omega),\label{eq:foddm_SDW}
\end{eqnarray}
Thus, in contrast to Sec.~\ref{sec:OPDW}, the even-frequency component of the PDW is now odd under $M$ (from Eq.~\eqref{eq:fevenm_SDW}). As the EPDW is even under $T$, to satisfy $SMT=-1$, $S$ should be even or in a spin-triplet configuration. Similarly, the odd-frequency component is even under $M$ (from Eq.~\eqref{eq:foddm_SDW}), odd under $T$, and thus $S$ should also be even or spin-triplet. So, both the even and odd-frequency components of the induced PDW are spin-triplet in nature when the SC coexists with the SDW.

It is actually not surprising that the PDW correlations in the coexistence state of the SC and the SDW are spin-triplet in nature. The coexistence or competition of the SC and the AFM was discussed within an SO(5) model in Ref.~[\onlinecite{Zhang97}]. Within this field theoretic picture, a PDW operator rotates a superconducting field to an AFM field. Since we consider the superconducting state to be spin-singlet, the spin transfer from the superconducting state to the AFM state is unity. Thus, the PDW operator has to be triplet with net spin unity, in order to conserve the spin. With $Q=\left(\pi,\pi\right)$, this PDW operator is famously known as the `$\Pi$' operator. PDW correlations, obtained in Eqs.~\eqref{eq:feven_SDW} and \eqref{eq:fodd_SDW}, are equivalent to the `$\Pi$' operator if we take $Q=\left(\pi,\pi\right)$. While the even frequency is discussed in the literature in the context of SO(5) model, the odd-frequency component has, to our knowledge, not been previously explored. 

Another striking difference between the OPDW generated in the two coexistence states is that, while the OPDW from CDW in Eq.~\eqref{eq:fodd} is zero when $\Delta_{k}=\Delta_{k+Q}$, the OPDW from SDW in Eq.~\eqref{eq:fodd_SDW} is maximum when $\Delta_{k}=\Delta_{k+Q}$ and zero when $\Delta_{k}=-\Delta_{k+Q}$. So, a coexistence of $d$-wave SC and SDW with $Q=\left(\pi,\pi\right)$ always lead to a zero OPDW. Thus, a coexistence of $d$-wave SC and antiferromagnetism in cuprates cannot induce any OPDW. On the other hand, spin-triplet OPDW correlations can exist in an $s$-wave superconductor, where the spin-singlet CDW-generated OPDW instead cannot be found. 

A coexistence regime of an $s$-wave SC and SDW with $Q=\left( \pi,0 \right)$ or $Q=\left( 0,\pi \right)$ is often observed in the iron-based superconductors (FeSC), especially in the ferropnictides \cite{Dai15}. Although FeSC are multi-orbital in nature, we can gain a simplistic understanding of the phenomenology by looking only at a single orbital described by the Hamiltonian in Eq.~\eqref{eq:Hamil_SDW}. FeSC are considered to be sign changing $s$-wave superconductors ($s^{+-}$ between the electron and the hole band) \cite{Hirschfeld11}, with the intra-orbital pair symmetry taking forms such as $s_{x^2y^2}\sim \cos(k_x)\cos(k_y)$ \cite{Seo08} with $\Delta_{k}=-\Delta_{k+Q}$. From Eqs.~\eqref{eq:fodd_SDW} we then directly see that $s_{x^2y^2}$ intra-orbital pairing cannot give rise to OPDW in this simplified one band picture. However, the SDW order has been shown to be able to promote $s^{++}$ intra-orbital pairing in the coexistence phase \cite{Hinojosa14}. In the $s^{++}$ state, induced spin-triplet OPDW are enhanced as $\Delta_{k}=\Delta_{k+Q}$. Furthermore, a transition from the $s^{+-}$ to the $s^{++}$ state is expected with the increase of impurity scatterings \cite{Golubov95,Efremov11,Efremov13}. This might then lead to the emergence of a spin-triplet OPDW with increasing disorder in FeSC. A complete understanding of the OPDW in the coexisting state of the SC and the SDW in the FeSC will require the consideration of multiple orbitals and is left to future work.  

\section{Conclusion and Discussion}\label{sec:Discussion}
In summary, we showed that the $d$-wave nature of the superconducting state in the cuprate high-temperature superconductors leads to induced odd-frequency pair density wave (OPDW) correlations in the region of the phase diagram where the SC coexists with a CDW. We considered several different CDW wave vectors relevant to the cuprates and showed that the existence of the OPDW is extremely robust to the choice of the wave vector and also to the variations in the band structure found between different families of cuprates. We find that the OPDW is often significant in magnitude and becomes even equal or larger than the even-frequency PDW (EPDW) near the nodal regions in momentum space. Moreover, keeping the SC and CDW order parameters unchanged, we find that breaking the $C_4$ lattice symmetry in the CDW wave vector with a uni-axial wave vector further enhances the OPDW amplitude. We also do not restrict ourselves to the cuprates, but also show that the OPDW can also be found in the coexistent state of SC and SDW in the iron-based superconductors.

Direct experimental detection of odd-frequency superconductivity has been challenging in the past. In our results, the induced PDW correlations do not directly influence the one-electron spectral function because the quasiparticle energy spectrum is not affected. So, one-electron experimental probes, such as angle-resolved photoemission spectroscopy (ARPES) or scanning tunneling spectroscopy (STS), will not detect the induced PDW. However, two-electron response functions, will have signatures of these PDW correlations \cite{Montiel16a,Montiel17,Garrido17,Boyack17,Morice18b,Dentelski20}, and even be able to distinguish odd-frequency components. For example, the imaginary part of the density response function $\chi^{\prime \prime}(q,\Omega)$ can characterize various bosonic excitations or correlations at different momentum ($q$) and energies or frequencies ($\Omega$) depending on the experimental probe. Note that there are two sets of momentum and frequencies, one internal ($k$ and $\omega$) and the other external ($q$ and $\Omega$). While calculating the response function, we integrate over internal frequencies only and arrive at $\chi^{\prime \prime}(q,\Omega)$ as a function of momentum and frequency of the external probe. The contribution of the OPDW to this response function comes with terms proportional to $F_{k}^{o}(i\omega){F_{k+q}^{o}}^{\dagger}(i\omega+\Omega)$. So, after integrating over $k$ and $\omega$, the OPDW contribution will still remain finite even though it is odd in frequency and would cancel  if only a single $F^o(i\omega)$ were to be integrated over frequencies. The same structure appears in other two-electron response functions, making them a good tool to probe odd-frequency correlations.

Now, the question remains how to distinguish between the EPDW and the OPDW contributions? To approach an answer to this question, we use the combined facts that the OPDW and the EPDW have different momentum space structure and different frequency dependence (EPDW peaks at $\omega=0$, while OPDW peaks at finite $\omega$). So, we should look for experimental probes which resolve the response functions both in momentum and frequency spaces. Probes like resonant inelastic x-ray scattering (RIXS), momentum-resolved electron-loss spectroscopy (M-EELS), and Raman spectroscopy can fulfill our needs. Each of these has their own advantages. 
RIXS is a well-established method in the literature, with RIXS experiments in underdoped BSCCO showing intensity peaks near $q=Q$ (where $Q$ is the CDW wave vector) both at $\Omega=0$ and finite $\Omega$ \cite{Chaix:2017fs}. In fact, the momentum-dispersion of the finite $\Omega$ peak has been predicted to be signatures of the role of phonons \cite{Chaix:2017fs} or collective modes \cite{Morice18b}. 
Due to the difference in the frequency dependence of the OPDW and the EPDW, we expect the frequency-dispersion to play a key role in detecting the OPDW. At the same time, RIXS lacks good energy resolution, where instead M-EELS derives a clear advantage in having a very high energy resolution of 1 meV compared to a resolution of 40 meV in RIXS \cite{Vig17}.  But, to the best of our knowledge, M-EELS have so far only been used to investigate optimal or overdoped cuprates \cite{Mitrano18}. Future M-EELS experiments in the underdoped regime are hence necessary. On the other hand, Raman spectroscopy has a unique advantage of preferentially probing parts of the Brillouin zone \cite{Devereaux97,Loret20}. For example, one way to distinguish the OPDW from the EPDW is Raman measurements in the B$_{2g}$ channel (which preferentially probes the nodal region), as both OPDW and EPDW are equally dominant in the nodal region. Moreover, the OPDW magnitude is also of the same order of magnitude as the uniform SC in the nodal region. Hence, it is possible that the so-called peak-dip-hump structure \cite{Eschrig06} in Raman intensity \cite{Loret20} as a function of frequency can find its explanation in terms of the OPDW. Finally, the OPDW might also leave distinct signatures in Josephson scanning tunneling measurements \cite{Hoshino16}, Meissner response \cite{Boyack17} and other transport measurements \cite{Tsvelik19}.

Let us also comment on disorder-induced inhomogeneities that are intrinsic to all cuprate materials. Due to the Imry-Ma criterion \cite{Imry75}, any strength of disorder disrupts the long-range phase coherence of CDW in dimensions $d\le 4$. Thus, at low or zero magnetic fields, CDW in two-dimensional cuprates, are only short-ranged correlations with no long-range phase coherence. We have ignored these fluctuations due to disorder in our mean-field calculations. However, the correlation lengths of CDW in cuprates are still long enough \cite{Ghiringhelli12} to allow for a finite CDW mean-field amplitude, validating the bulk of our analysis in this work. Even if disorder fluctuations were to be included on top of the mean-field analysis, we expect the OPDW to be present locally in regions of local coexistence of SC and CDW. In addition, the high-field CDW is experimentally shown to be a `true' long-range order \cite{Gerber:2015gx,Chang16,LeBoeuf13}. The high-field coexistence of SC and uni-axial CDW will thus lead to a long-range OPDW.    

We have also not explored the prospects of the OPDW being responsible for the anomalous properties of the pseudo-gap phase in cuprates. Our results should motivate future research in this direction as its dynamical character naturally make it a hard to detect, or even hidden, order. Moreover, in the pseudo-gap phase, there are additional broken symmetry orders, such as nematic or time-reversal symmetry broken orders. As we found in Sec.~\ref{sec:uniaxial}, the magnitude of the OPDW is significantly increased when the $C_4$ symmetry is broken by the CDW wave vectors. We believe an additional nematic distortion of the Fermi surface will likely  further enhance the OPDW correlations. 

We finally note two possible extensions. First, the translational symmetry breaking due to the CDW order reconstructs a single band superconductor into effectively two bands, one shifted from the other by the CDW wave vector. This shift makes the system conceptually analogous to a multiband superconductor with different superconducting order parameters in different bands. As a result, odd-frequency correlations are induced in the same spirit as in multiband superconductors, but with a major difference being the modulations in the induced order. As a consequence, our work on coexisting CDW and SC can easily be generalized to other translational symmetry breaking orders, which might thus also host significant odd-frequency components. 
Second, in the one-band model considered in this work, the coexistence of SC and CDW cannot give rise to an odd-frequency component of the induced PDW when the superconducting order parameter is momentum-independent or $s$-wave. But the analysis does not hold true for multi-band superconductors, such as transition-metal dichalcogenides (TMDs) with coexistence of SC and CDW. A PDW state in TMDs has already been proposed theoretically \cite{Venderley19} and also very recently observed experimentally \cite{Liu20}. Even though TMDs host $s$-wave SC, the multiband nature might still induce OPDW. The search for OPDW in TMDs is a part of ongoing research.    

\begin{acknowledgments}

We thank J.~Cayao for useful discussions. We gratefully acknowledge  financial support from the Swedish Research Council (Vetenskapsr\aa det, Grant No.~2018-03488), the Knut and Alice Wallenberg Foundation through the Wallenberg Academy Fellows program, and the European Research Council (ERC) under the European Unions Horizon 2020 research and innovation programme (ERC-2017-StG-757553).

\end{acknowledgments}

 \bibliographystyle{apsrev4-1}
\bibliography{Cuprates}

\end{document}